\documentclass[12pt]{article}

\usepackage{xspace}
\usepackage{epsfig}

\newcommand{\A}{{\cal A}}
\newcommand{\B}{{\cal B}}
\newcommand{\F}{{\cal F}}
\newcommand{\W}{{\cal W}}
\newcommand{\Z}{{\cal Z}}

\newcommand{\nc}{\newcommand}

\def\simgt{\stackrel{>}{{}_\sim}}

\nc{\beq}{\begin{equation}}
\nc{\eeq}{\end{equation}}
\nc{\beqa}{\begin{eqnarray}}
\nc{\eeqa}{\end{eqnarray}}
\nc{\bea}{\begin{eqnarray}}
\nc{\eea}{\end{eqnarray}}
\nc{\ra}{\rightarrow}
\nc{\lsim}{\begin{array}{c}\,\sim\vspace{-21pt}\\< \end{array}}
\nc{\gsim}{\begin{array}{c}\sim\vspace{-21pt}\\> \end{array}}

\nc{\LL}{L}
\nc{\vv}{\tilde{v}}

\title{
\vspace*{-1.3cm}
\begin{flushright}
\normalsize{
ANL-HEP-PR-02-115\\
EFI-02-43\\
FERMILAB-PUB-02/342-T
  }
\end{flushright}
\vspace{1.5cm}
\Large
\textbf{ Opaque Branes in Warped Backgrounds
} 
\vspace*{1.0cm}
\author{\large\textbf{Marcela Carena$^a$}, \textbf{Eduardo Pont\'{o}n$^a$},\\[0.3cm]  
\textbf{Tim M.P. Tait$^a$},
and \textbf{C.E.M.~Wagner$^{b,c}$}\\ \\[0.5cm]
$^a$\normalsize\emph{Fermi National Accelerator Laboratory,
P.O. Box 500, Batavia, IL 60510, USA} \\
$^b$\normalsize\emph{HEP Division, Argonne National Laboratory,
9700 Cass Ave.,
Argonne, IL 60439, USA} \\
$^c$\normalsize\emph{Enrico Fermi Institute, Univ. of Chicago, 5640
Ellis Ave., Chicago, IL 60637, USA}}}

\begin{document}
\setcounter{page}{0}
\maketitle
\vspace*{1cm}
\begin{abstract}
We examine localized kinetic terms for gauge fields which can 
propagate into compact, warped extra dimensions.  We show that these 
terms can have a relevant impact on the values of the Kaluza-Klein 
(KK) gauge field masses, wave functions, and couplings to brane and 
bulk matter.  The resulting phenomenological implications are 
discussed.  In particular, we show that the presence of opaque branes, 
with non-vanishing brane-localized gauge kinetic terms, allow much 
lower values of the lightest KK mode than in the case of transparent 
branes.  Moreover, we show that if the large discrepancies among the 
different determinations of the weak mixing angle would be solved in 
favor of the value obtained from the lepton asymmetries, bulk 
electroweak gauge fields in warped-extra dimensions may lead to an 
improvement of the agreement of the fit to the electroweak precision 
data for a Higgs mass of the order of the weak scale and a mass of the 
first KK gauge boson excitation of a few TeV, most likely within reach
of the LHC.
\end{abstract}

\thispagestyle{empty}
\newpage

\setcounter{page}{1}

\baselineskip18pt

\section{Introduction}

Much of the recent interest in theories with extra dimensions stems 
from the fact that they provide a possible solution to the hierarchy 
problem of the Standard Model.  In the case of flat extra dimensions 
\cite{Arkani-Hamed:1998rs}, the fundamental Planck scale could take 
values of order of the weak scale.  The gravity interactions at long 
distances are governed by the four dimensional Planck scale, $M_{Pl}$, 
which is related to the fundamental Planck scale $M$ by a factor 
proportional to the volume of the extra dimensional space.  Therefore, 
for large values of the compactification scale and/or large number of 
extra dimensions, one can reconcile the observed value of the Planck 
scale with a fundamental scale $M \simeq {\cal{O}}$(TeV).

Since this mechanism demands an a priori unexplained large 
extra-dimensional volume, it could be argued that rather than 
providing a solution to the hierarchy problem, theories with flat 
extra dimensions allow a reformulation of the problem.  On the other 
hand, flat extra dimensions would only be possible in the presence of 
a tension-less brane, confining all SM fields.  Under these 
conditions, however, it is somewhat more natural to think about a 
brane with finite tension, and an induced curvature in the extra 
dimensional scenarios.

Non-vanishing curvature provides a distortion of the metric, which 
allows new alternatives to address the hierarchy problem.  For 
instance, one can find simple solutions to the case of branes with 
finite tension by assuming that the local four dimensional metric is 
affected by an exponential warp factor, which depends linearly on the 
position of the brane.  In this simple famework, one can get a 
solution to the hierarchy problem without assuming any unnatural large 
factor \cite{Randall:1999ee}.  Indeed, assuming all fundamental mass 
scales to be of the same order, the ratio of the physical Higgs vacuum 
expectation value to the observable four dimensional Planck scale is 
exponentially suppressed, with an exponent that depends on the warp 
factor $k$ times the position $L$ of the Higgs brane in the extra 
dimensions.  Taking $k L \simeq 34.5$ provides a good solution to the 
hierarchy problem.  It is important to emphasize that, as in any other 
solution, new physics appears at the TeV scale.  This new physics
includes graviton Kaluza Klein states and a graviscalar, also called 
the radion, with effective interactions suppressed by a scale of order 
of the TeV, $M e^{-k L}$.

It is natural to assume that not only the Higgs field, but all matter 
fields are confined to the brane at $y=L$, which we shall call the 
infrared brane.  The gauge fields, however, may propagate in the extra 
dimensions.  One motivation for this is the fact that, from the field 
theoretical point of view, it is more difficult to localize gauge 
fields than to localize fermions.  However, there is a 
phenomenological obstacle to this realization : the Kaluza Klein
excitations of the gauge bosons couple with a strength $\sqrt{2\, k \, 
L}$ times the zero mode coupling, an order of magnitude larger for 
values of $k \, L$ necessary to provide a solution of the hierarchy 
problem.  The exchange of these strongly coupled higher KK states can 
be used to set a lower bound on the mass of the first KK excitation 
that is about 20 TeV \cite{Davoudiasl:1999tf}.

Precision electroweak observables present additional challenges, since
these observables are affected at the tree-level
\cite{Csaki:2002gy,Burdman:2002gr,Huber:2000fh}.
The modification of these observables is related to the shape of the 
extra dimensional wave functions in the presence of a localized 
symmety breaking VEV. The presence of the localized scalar VEV induces 
a repulsion of the gauge fields from the brane, leading to a breakdown 
of the usual linear relation between the gauge boson mass with its 
coupling and the Higgs VEV. In order to recover acceptable 
phenomenological predictions, the bound on the first KK excitation has 
to be pushed to even larger values than the ones obtained from the KK 
boson exchange.

In this article we analyze how the results for the gauge field 
interactions described above is modified by the presence of local 
gauge field kinetic terms.  Such terms are naturally expected to be 
present in any realistic theory, and indeed will be induced 
radiatively by the localized Higgs and fermions on the IR brane even 
if the underlying dynamics is such that they vanish at any particular 
scale \cite{Dvali:2000rx, Georgi:2000ks, Carena:2002me}.  The presence of a local
brane kinetic term on the infrared brane implies that at sufficiently
high energies, the gauge interactions on the brane should be four 
dimensional and renders the brane opaque to gauge fields of short 
wavelength along the extra dimension.  This can only be possible if 
there is an important modification of the couplings of the gauge KK 
modes to matter localized on the brane 
\cite{Carena:2002me,Carena:2001xy,Dvali:2001gm}.  Moreover, for 
sufficiently large local gauge kinetic terms, the physics on the brane 
should be dominated by the four dimensional behavior, and therefore it 
should be possible to relax these bounds on the mass of the lightest 
gauge boson KK mass.

The plan of this work is as follows.  In section~\ref{sec:prop}, we 
shall give the expression for the gauge field propagator with brane 
kinetic terms.  We shall analyze its behavior in different momentum 
regimes and for different values of the local gauge kinetic terms.  In 
section~\ref{sec:KK}, we proceed with the Kaluza-Klein decomposition.  
After reviewing the situation in the case of transparent branes, in 
the absence of local gauge kinetic terms, we analyze the case of one 
or two opaque branes.  We also provide a comparison of the results 
obtained with the KK decomposition with the ones derived from the 
behavior of the five dimensional gauge field propagator.  In 
section~\ref{sec:higgs}, we discuss the effects induced by the 
presence of a localized Higgs VEV. We discuss both the case of 
vanishing local gauge kinetic terms and the effects associated with 
the presence of these local terms.  In section~\ref{sec:pheno} we 
apply our results to the case where the electroweak gauge bosons 
propagate in the bulk, and analyze the phenomenological consequences 
in this scenario.  In particular, we show that the presence of the 
brane-localized gauge kinetic terms opens the possibility for the 
first KK excitations to be within reach of the LHC. We reserve 
section~\ref{sec:concl} for the conclusions.

\section{Gauge Field Propagator with Brane Kinetic Terms}
\label{sec:prop}
In this section we derive the tree-level propagator for a gauge field 
described by the action
\beq
\label{action}
S = -\frac{1}{4 g_5^2} \int d^4xdy \sqrt{-g} \left[ {\cal F}^{MN} 
{\cal F}_{MN}
 + 2 \,\delta(y) r_{UV} {\cal F}^{\mu\nu} {\cal F}_{\mu\nu} 
 + 2 \,\delta(y-L) r_{IR} {\cal F} ^{\mu\nu} {\cal F}_{\mu\nu} \right]~,
\eeq
where $g$ is the determinant of the metric, and capital latin letters 
refer to the full 5d coordinates, $M = 0, 1,2 ,3, 5$, whereas lower 
case greek letters refer only to the four uncompactified dimensions, 
$\mu = 0, 1, 2, 3$.  In the above we have rescaled the bulk kinetic 
term by absorbing $g_5$ into ${\cal A}_M$, so that ${\cal A}_M$ has 
mass dimension one, the canonical dimension for a gauge boson 
propagating in four dimensions.  Then $g_{5}^{-2}$ has dimensions of 
mass and the coefficients of the local brane terms, $r_i = g_5^2 / 
g_i^2$, have dimensions of length.\footnote{For future convenience, we 
have factored out an explicit factor of two in the definition of the 
localized terms.  The physical fifth dimension corresponds to the 
interval $[0,L]$, with the branes located at the endpoints.  
Consequently, each delta function contributes a factor of 1/2 when 
performing the $y$ integration.} ${\cal F}$ is the usual 
field-strength functional of the gauge fields, \bea {\cal F}^a_{M N} 
&=& \partial_M {\cal A}^a_N - \partial_N {\cal A}^a_M + f^{abc} {\cal 
A}^b_M {\cal A}^c_N , \eea for a non-Abelian Yang-Mills theory, with 
the final term omitted in the Abelian case.  For simplicity of 
notation, we denote the extra dimensional coordinate $x_5$ as $y$.  
The full set of five dimensional coordinates are denoted by capital 
letters, i.e.  $X = \{ x^\mu, y \}$.


We assume a background metric defined by the line element,
\beq
\label{lineelement}
ds^{2} = e^{-2\sigma} \eta_{\mu\nu} dx^\mu dx^\nu + dy^2 ,
\eeq
with $\eta_{{\mu\nu}} = \mathit{diag}(-1,+1,+1,+1)$, $\sigma(y) = k 
|y|$ and $0\leq y \leq L$.  The interval is assumed to arise from a 
$Z_{2}$ orbifold such that ${\cal A}_{\mu}$ is even and $\A_5$ is odd 
under reflection in $y$.  It is then always possible to choose a gauge 
in which $\A_5=0$, corresponding to a unitary gauge.  This will be 
sufficient for our purposes, but it is also interesting to exhibit the 
full gauge-dependence in an $R_\xi$ gauge.

In an $R_\xi$ gauge both the four dimensional vector field $\A_\mu$
and the scalar $\A_5$ propagate.  The gauge choice is defined such 
that terms which mix the two are zero.  This is enforced by the 
gauge-fixing term,
\bea
-\frac{1}{2 \xi g_5^2} \left[ \partial^\mu \A_\mu 
- \xi \partial_y \left( e^{-2ky} \A_5 \right) \right]^2 .
\eea
We could work out the ghost interactions corresponding to this choice 
of gauge-fixing, but will not need them for our (tree-level) analyses.  
The propagators $G_{{\mu\nu}}(X,X') = \langle A_{\mu}(X) A_{\nu}(X') 
\rangle$ and $G_{{55}}(X,X') = \langle A_{5}(X) A_{5}(X') \rangle$, 
satisfy
\bea
\frac{1}{g_5^2} \left\{ P^{\mu \nu} +
\eta^{\mu\nu} \partial_y [e^{-2\sigma} \partial_y] +
2 \, r_i \delta(y-y_i) P^{\mu \nu} 
+ \frac{1}{\xi} \partial^\mu \partial^\nu \right\}
G_{\nu\alpha}(X,X') & = & \delta^\mu_\alpha \delta(X-X')\\
\frac{1}{g_5^2} \left\{ \partial^2 \; + \; \xi \, \partial_y^2 \, e^{-2ky}
\right\} G_{55}(X,X') & = & \delta(X-X')~,
\eea
where $P^{\mu \nu} \equiv \eta^{\mu \nu} \partial^2 - \partial^\mu 
\partial^\nu$ and $r_i \delta(y-y_i)$ indicates the sum over both the 
UV and IR branes.  

Due to the translational invariance of Eq.~(\ref{lineelement}) along 
the four noncompact dimensions, it is convenient to work in mixed 
position and momentum space, defined by $G(p;y,y') = \int d^{4}\!x
e^{i \eta_{\mu\nu} p^{\mu} x^{\nu}} G(x,y,y')$.  The momentum defined 
in this way is a conserved quantity.  However, it is important to keep 
in mind that it is not necessarily the physical momentum of the 
propagating field.  Rather, from Eq.~(\ref{lineelement}) it can be 
inferred that the momentum that an observer would measure when 
standing at $y$ is $p_{phys} = e^{k y} p$.  In the following, we will 
be working with the ``coordinate'' momentum $p$, as is standard in the 
literature, with the understanding that $p$ is the momentum as 
measured by UV observers.  Keeping the fifth coordinate explicit will 
also make the locality properties in $y$ manifest.  We choose to work 
in Euclidean space, which will simplify the analysis later.  Then the 
$G_{\mu\nu}(p;y,y')$ propagator can be written as
\beq
G_{\mu\nu}(p;y,y') = \left(\eta_{\mu\nu} - \frac{p_\mu p_\nu}{p^2} \right) 
G_p(y,y') + \frac{p_\mu p_\nu}{p^2} G_p^\prime (y,y')~,
\eeq
where $G_p(y,y')$ and $G^\prime_p(y,y')$ satisfy
\bea
\frac{1}{g_5^2} \left\{ \partial_y [e^{-2\sigma} \partial_y] 
- p^2 [1+2 \, r_i \delta(y-y_i)] \right\}
G_p(y,y') = \delta(y-y')~, \\
\frac{1}{g_5^2} \left\{ \partial_y [e^{-2\sigma} \partial_y] 
- \frac{p^2}{\xi}  \right\} G^\prime_p(y,y') = \delta(y-y')
~.
\eea
The $\delta$-functions impose the following boundary conditions at
$y = 0, \LL$:
\beqa
\label{propbound0}
\left[ \partial_y G_p -
r_{UV} p^2 e^{2\sigma} G_p \right]_{y = 0} &=& 0   \\
\label{propboundL}
\left[ \partial_y G_p +
r_{IR} p^2 e^{2\sigma} G_p \right]_{y = \LL} &=& 0~,
\eeqa
whereas $\partial_y G^\prime_p$ must vanish at the boundaries.  At
$y = y'$, both $G_p$ and $G^\prime_p$ must be continuous and satisfy
\beq
e^{-2 \sigma} \left[ \partial_y G^{(\prime)}_p |_{y=y'+\epsilon} 
- \partial_y G^{(\prime)}_p |_{y=y'-\epsilon} \right]
= g_{5}^{2}~,
\eeq
where $\epsilon \rightarrow 0^{+}$. The solutions can be written as
\beq
\label{propagator}
G_p(y,y') = \frac{g_{5}^{2}e^{k(y+y')}}{k(AD - BC)}
\left[ A K_1\left( \frac{p}{k} e^{k y_<} \right) +
B I_1\left( \frac{p}{k} e^{k y_<} \right) \right]
\left[ C K_1\left( \frac{p}{k} e^{k y_>} \right) +
D I_1\left( \frac{p}{k} e^{k y_>} \right) \right]
\eeq
where $K_{\alpha}$, $I_{\alpha}$ are modified Bessel functions of 
order $\alpha$, $y_{<(>)}$ are the smallest (largest) of $y$, $y'$ and
\beqa
\label{coefficients}
A &=& I_0\left(\frac{p}{k}\right) -
p \, r_{UV} I_1\left(\frac{p}{k}\right) \nonumber \\
B &=& K_0\left(\frac{p}{k}\right) +
p  \, r_{UV} K_1\left(\frac{p}{k}\right) \nonumber \\
C &=& I_0\left(\frac{p}{k}e^{k L}\right) +
p  \, e^{k L} r_{IR} I_1\left(\frac{p}{k}e^{k L}\right) \\
D &=& K_0\left(\frac{p}{k}e^{k L}\right) -
p  \, e^{k L} r_{IR} K_1\left(\frac{p}{k}e^{k L}\right)~. \nonumber
\eeqa
The solution for $G^\prime_p$ takes the same form with $p \ra p / \xi$ 
and $r_{UV}=r_{IR}=0$ in Eq.~(\ref{coefficients}).
The $G_{55}(p; y,y^\prime)$ propagator is of the same form as 
$G^\prime_p$, but with second order Bessel functions and multiplied by 
$1/\xi$.  From these results we see how the transition into the 
unitary gauge, $\xi \ra \infty$ (or $\A_5 = 0$) proceeds; $G_{55} \ra 
0$ and $G^\prime_p \ra G_0$.  In this gauge, all of the physics is 
contained in Eqs.~(\ref{propagator}) and (\ref{coefficients}).

To understand the physics contained in Eq.~(\ref{propagator}) it is 
instructive to analyze it in various limits.  We define the following 
auxiliary function
\beq
B(Y_>,Y_<,r) = \frac{1}{1 + p \, r}
\left\{ \left[ 1 + e^{-\frac{2p}{k}(e^{k Y_>} - e^{k Y_<})} \right] +
p \, r \left[ 1 - e^{-\frac{2p}{k}(e^{k Y_>} - e^{k Y_<})} \right]
\right\}~,
\eeq
which, as the notation indicates, will be used only when $Y_> > Y_<$.  
In this case, $B(Y_>,Y_<,r)$ has a very simple behavior when the 
condition $(2p/k) e^{k Y_>} \gg 1$ is satisfied: $B \approx 2/(1 + p 
\, r)$ when $Y_<$ is (extremely) close to $Y_>$, and $B \approx 1$ 
otherwise.  Specializing to the case of interest for addressing the 
hierarchy problem, where $e^{-k L} \ll 1$, we can write:

i) For $p \gg k$,
\beqa
\label{veryhighEpropagator}
G_p(y,y') &\sim& -\frac{g_5^2}{2p}\,
e^{\frac{1}{2}k(y+y')}
e^{-\frac{p}{k}( e^{k y_>} - e^{k y_<})}
B(y_<,0,r_{UV}) B(L,y_>, e^{k L} r_{IR})~.
\eeqa

ii) For $k \gg p \gg k e^{-k y_<}$,
\beqa
\label{highEpropagator}
G_p(y,y') &\sim& -\frac{g_5^2}{2p}\,
e^{\frac{1}{2}k(y+y')}
e^{-\frac{p}{k}( e^{k y_>} - e^{k y_<})} B(L,y_>, e^{k L} r_{IR})~.
\eeqa

iii) For $k e^{-k y_<} \gg p \gg k e^{-k y_>}$,
\beqa
\label{nonlocalpropagator}
G_p(y,y') &\sim& - \frac{g_5^2}{2 p^2}\,
e^{\frac{1}{2}k y_>} e^{-\frac{p}{k} e^{k y_>}}
\sqrt{\frac{\pi p}{2 k}} \; \frac{ 2 k + (e^{2 k y_<}-1) p^2 r_{UV} }
{\ln(2k/p) + k r_{UV}}
\, B(L,y_>, e^{k L} r_{IR})~.
\eeqa

iv) For $k e^{-k y_>} \gg p \gg k e^{-k L}$,
\beq
\label{lowEpropagator}
G_p(y,y') \sim - \frac{g_5^2}{2 p^2}\;
\frac{2 k+(e^{2 k y_<}-1) p^2 r_{UV}}
{\ln(2k/p) + k r_{UV}}~.
\eeq

v) For $p \ll k e^{-k L}$,
\beq
\label{verylowEpropagator}
G_p(y,y') \sim - \frac{g_5^2}{2 k p^2}\,
\frac{[2 k+(e^{2 k y_<}-1) p^2 r_{UV}]
[2 k+(e^{2 k L}-e^{2 k y_>}) p^2 r_{IR}]}
{2 k (L + r_{UV} + r_{IR}) + e^{2k L} p^2 r_{UV} r_{IR}}~.
\eeq

Several remarks are in order.  Consider first the properties of the 
propagator when the kinetic terms vanish.  If we are doing physics at 
a fixed location along the warped dimension (say, on one of the 
branes), it follows from the previous asymptotic forms that the 
propagator $G_p(y,y) $ changes from a four dimensional ($\sim 
1/p^{2}$) to a five dimensional ($\sim 1/p$) behavior at a scale $p 
\sim k \, e^{-k y}$.  The onset of the higher dimensional scaling 
indicates that the gauge theory ceases to be predictive when the 
external 4d momenta are much larger than $k \, e^{-k y}$, and that it 
should be cut off at a scale $\Lambda \, e^{-k y}$, for some constant 
$\Lambda \gsim k$.  Thus, a $y$-dependent cutoff on $p$ appears 
naturally in this language.{\footnote{The $y$-dependence of the cutoff 
is simply a consequence of general covariance and the fact that AdS is 
homogeneous.  The statement that the highest energy modes described by 
the effective theory have $p \sim \Lambda \, e^{-k y}$ translates in 
position space into minimum {\it coordinate} wavelengths of size 
$\Delta x \sim 1/p \sim e^{k y}/\Lambda$.  That is to say, the 
effective theory describes {\it proper} wavelengths larger than 
$\Delta s = e^{-k y} \Delta x \sim 1/\Lambda$, independent of the 
position in the bulk.  By a similar reasoning, one can see that the 
powers of the warp factors $e^{k y}$ in 
Eqs.~(\ref{veryhighEpropagator})--(\ref{verylowEpropagator}) are also 
dictated by general covariance.}}

For the consistency of this picture, it is important to note that, at 
energy scales above $k \, e^{-k y}$, the observables localized at $y$ 
are only sensitive to the physics of nearby points $y'$, such that 
$\exp(- p |e^{k y} - e^{k y'}|/k) \sim 1$.  Furthermore, case iii) 
shows that the contributions from far away points $y'>y$ are 
exponentially suppressed for all momenta above the lower cutoff
$\Lambda\,e^{-k y'}$.  Thus, the physics associated with some energy
scale at $y$ is effectively shielded from contributions that would be
outside the region of validity of the effective theory.

It is remarkable that the gauge field tree-level propagator with 
endpoints at $y$ exhibits a purely four-dimensional scaling at all 
scales below $k \, e^{-k y}$.  This property together with the 
shielding discussed in the previous paragraph are an important 
ingredient in the understanding of the four-dimensional ``running" of
the gauge couplings that has been discussed recently 
\cite{gaugerunning,Goldberger:2002cz}.  In fact, if one defines the 
gauge coupling at $y$ at a scale $p$ via the two point gauge 
correlator with endpoints at $y$ and external momentum $p$, as 
advocated in \cite{Goldberger:2002cz}, one can understand from the 
above discussion that, provided $p < k \, e^{-k y}$, all the 
propagators that are relevant in a loop calculation will exhibit a 
four dimensional scaling.  Therefore, such momentum integrals will 
depend logarithmically on $p$, with no power-law sensitivity.

From the behavior of the five dimensional propagator $G_{p}(y,y')$, 
Eqs.~(\ref{veryhighEpropagator})--(\ref{verylowEpropagator}), we can 
also quickly understand some of the effects induced by the presence of 
the local terms.  Considering again coincident points $y = y'$, we see 
that for $p \gg k \, e^{-k y}$, there are three different regions.  
Away from the boundaries we find, as before, the five dimensional 
scaling $G_{p}(y,y) \sim 1/p$, and we conclude that the brane kinetic 
terms have an exponentially small effect (that does not show up at 
this order), as expected from locality.  On the branes, however, we 
find $G_{p}(0,0) \sim g_5^2/(p+p^{2} r_{UV})$ and similarly for 
$G_{p}(L,L)$ with $e^{k L} r_{IR}$ in place of $r_{UV}$ and $g_5^2 
e^{kL}$ in place of $g_5^2$.  This shows that, for brane observers, a 
four dimensional behavior is recovered for $p \gg 1/r_{c}$ where 
$r_{c} = r_{UV}$ on the UV brane and $r_{c} = e^{k L} r_{IR}$ on the 
IR brane.  Note that the IR crossover distance has been appropriately 
red-shifted as is generally expected for dimensionful parameters in 
the Randall-Sundrum setup.

We postpone a further analysis of the physics behind the behavior of 
the propagator when $r_{UV}$ and $r_{IR}$ are nonzero to the next 
section.  Here we only note that, from a theoretical point of view, it 
is perfectly consistent to take the coefficients $r_{UV}$, $r_{IR}$ to 
be large in units of the fundamental length $1/M$, as long as we 
include their effects exactly as has been done here.  As we have seen, 
the previous exact propagator does not change the divergence structure 
of the 5d theory, except on the branes, where it actually softens it.  
In particular, it is consistent to assume that all other couplings are 
small so that perturbation theory is valid, even if the brane kinetic 
terms are larger than their ``natural'' scale \cite{Ponton:2001hq}.  
Nevertheless, when the gauge fields are identified with the carriers 
of the standard model interactions, one can derive some constraints on 
the size of the brane kinetic terms from the fact that the observed 
standard model gauge couplings are of order one.  For simplicity, and 
because it is the case that will interest us in the phenomenological 
applications of section 5, let us consider the case where $r_{UV}=0$.  
From Eq.~(\ref{verylowEpropagator}), we see that the observed zero 
mode gauge coupling, $g_{0}$, is given by
\beq
\label{4Dgaugecoupling}
g_{0}^{2} = \frac{g_{5}^{2}}{\LL + r_{IR}}
=\frac{g_{5}^{2}}{\LL} \left(1 + \frac{r_{IR}}{\LL} \right)^{-1}~.
\eeq
In the limit that $r_{IR} \gg L$, this gives $r_{IR} \simeq g_5^2/g_0^2$.
This relation does not directly impose a bound on $r_{IR}/\LL$ which,
as will be shown in the following sections is the most relevant quantity.
However, if one assumes that the gauge observables at any given position
$y$ are perturbative below the local ``compactification scale" $k\, e^{-k y}$,
as would be desirable from the point of view of gauge coupling unification,
one has $g_5^2 k \lsim 16 \pi^2$, a four-dimensional loop factor (this can
be inferred from Eq.~(\ref{lowEpropagator})). Under this assumption, it
follows that $r_{IR}/\LL \lsim 16 \pi^2/( g_0^2k \LL)$. Recalling now
that a solution to the hierarchy problem requires $k \LL \sim 35$, we
conclude that $r_{IR}/\LL \lsim 5/g_0^2$, which can be of order twenty for
the electroweak interactions.

\section{Kaluza-Klein Decomposition}
\label{sec:KK}

From the point of view of an observer at low energies, the relevant 
description of the phenomenology is in terms of the Kaluza-Klein 
modes.  Bulk gauge fields in warped backgrounds with transparent 
branes were considered in \cite{Pomarol:1999ad}.  In order to 
elucidate the effects of the opaque branes, we begin with a brief 
review of some of the features of the transparent brane case.

\subsection{Review of the Transparent Brane Case}

We begin with the linearized 5d equation of motion for a bulk gauge
field.  From the action, Eq.~(\ref{action}), with $r_i=0$, we derive
the equation for the $\lambda$ component of $A$,
\bea
\partial_\sigma \partial^\sigma \A^\lambda
- \partial^\lambda ( \partial_\sigma \A^\sigma )
- 2 k  e^{-2 k y} ( \partial_y \A^\lambda )
+ e^{-2 k y} ( \partial_y^2  \A^\lambda ) & = & 0 .
\eea
In order to determine the KK spectrum, we expand $\A(X)$ in wave
functions,
\bea
\A^\lambda (X) & = & \sum_n f_n(y) A_n^\lambda(x^\mu),
\eea
and require the $A_n^\lambda$ to obey the 4d equation of motion for a
free massive gauge field,
\bea
\partial_\sigma \partial^\sigma A_n^\lambda
- \partial^\lambda ( \partial_\sigma A^\sigma_n )
- m_n^2 A^\lambda_n & = & 0.
\eea
This requires the wave functions to obey the equation,
\bea
\left[\partial_y^2 - 2 k  \partial_y + e^{2ky} m_n^2 \right] f_n(y)
& = & 0 .
\eea
The solutions to this equation are Bessel functions,
\bea
f_n(y) & = & {\cal N}_m e^{k |y|} \left\{
J_1 \left( \frac{m_n}{k} e^{k|y|} \right)
+ b Y_1 \left( \frac{m_n}{k} e^{k|y|} \right)
\right\},
\label{eq:wavefunction}
\eea
where ${\cal N}_m$ is an overall normalization factor, determined by
requiring each KK mode to have canonically normalized kinetic terms,
\bea
\label{orto}
\frac{1}{g_5^2} \int_{0}^{L} dy \:
f_n(y) f_m(y) & = & \delta_{nm}, \nonumber \\
\frac{1}{g_5^2} \int_{0}^{L} dy \: e^{-2ky} \; f^\prime_n (y) f^\prime_m (y)
& = & m_n^2 \delta_{nm} .
\eea
and $b$ is a function of $m_n$, both of which are determined by boundary
conditions below.

In order to find the masses, we impose boundary conditions that the
first derivative of the wave functions be continuous at the $y=0$ and
$y= L$ boundaries.  This requires,
\bea
b^0 =  - \frac{J_0 \left( \frac{m_{n}}{k} \right)}
{Y_0 \left( \frac{m_{n}}{k} \right)} & , &
b^L =  - \frac{J_0 \left( \frac{m_{n}}{k} e^{k L} \right)}
{Y_0 \left( \frac{m_{n}}{k} e^{k L} \right)} .
\eea
Requiring both boundary conditions be satisfied simultaneously,
$b^0=b^L$, provides a transcendental equation for the quantized
masses.  This equation may be solved numerically, resulting in masses,
in units of $k \, e^{-kL}$, of $0$, and approximately $2.5$, $5.6$,
$8.7$,\ldots  \cite{Pomarol:1999ad}, where the difference between two
subsequent mass eigenvalues tends asymptotically to $\pi$.

The coupling of the $n$-$th$ KK gauge boson mass eigenstate to matter 
on the brane at $y=L$ is equal to $f_n(L)$ times the appropriate 
charge of the matter field.  For large values of $kL$, the KK modes 
couple universally (up to a sign) to fermions located on the IR brane 
at $y =L$, with a coupling $\sqrt{2\,k\,L}$ larger than the zero 
mode's coupling (for $kL \simeq 35$, the coupling is therefore 
approximately $8.5$ times larger than the zero mode coupling).  The KK 
modes couple to fields on the UV brane at $y = 0$ with couplings that 
vary with the KK mode number, and are typically suppressed compared to 
the zero mode coupling.  This behavior may be seen in Fig.  1, for 
$r_{IR} = 0$.

It is interesting to compare these results with the behavior of the
five-dimensional propagator for two points located on either the IR or
the UV brane.  For $p \gg k e^{-kL}$, that is for momenta much
larger than the mass of the lightest KK mode, the five dimensional
propagator with endpoints on the IR brane reads
\bea
\label{gpll}
G_p(L,L) & = & -\frac{g_5^2}{p} \; e^{k L}
\nonumber\\
& = & - \frac{2 \, g_5^2 \, k}{p^2} \; \frac{\pi}{2}
\left(\frac{p}{\pi \, k \, e^{-k L}}\right).
\eea
Using the result that, for $p \gg \tilde{m} \equiv \pi k e^{-k L}$,
one can approximate $ \sum_n [1/(p^2 + (n \tilde{m})^2)] \simeq (\pi/2
p^2) (p/\tilde{m})$, Eq.~(\ref{gpll}) has a clear interpretation.  The
five dimensional behavior of the propagator is given by the sum over
KK modes with a constant coupling $g_n^2 = 2 g_5^2 k$ to the IR brane.
Since the zero mode coupling is given by $g_0^2 = g_5^2/L$, the
relation between these couplings is the one described above.

On the UV brane, the propagator reaches a five dimensional behavior
only at momenta much larger than $k$.  Indeed, from
Eq.~(\ref{veryhighEpropagator}) one obtains
\bea
\label{gp00}
G_p(0,0) & = & - \frac{g_5^2}{p}.
\eea
The comparison between Eqs.~(\ref{gpll}) and (\ref{gp00}) suggests
that those KK modes with masses larger than $k$ couple with a constant
coupling $2 g_5^2 k e^{-k L}$ to the UV brane.  Observe that these
couplings are exponentially suppressed with respect to the couplings
on the IR brane.  A numerical evaluation of the couplings of
asymptotically large mass eigenstates for $kL \simeq 5$ confirms that
this is indeed the case.

\subsection{The Opaque Brane Case}

Now we allow for opacity on both the IR and the UV branes, allowing
$r_{IR} \not = 0$, $r_{UV} \not = 0$ in Eq.~(\ref{action}).  In this
case, the orthonormality conditions for the KK decomposition become,
\bea
\label{ortor}
\frac{1}{g_5^2} \int_{0}^{L} dy \:
\left[ 1 + 2 \, r_{UV} \delta (y) + 2 \, r_{IR} \delta (y-L) \right]
f_n(y) f_m(y) & = & \delta_{nm}, \nonumber \\
\frac{1}{g_5^2} \int_{0}^{L} dy \: e^{-2ky} \; f^\prime_n (y) f^\prime_m (y)
& = & m_n^2 \delta_{nm} .
\eea
Eq.~(\ref{ortor}) reduces to Eq.~(\ref{orto}) for $r_{UV} = r_{IR} =
0$.  These conditions may be simultaneously solved, as before, by
imposing the condition that the 4d gauge fields $A^\lambda_n(x^\mu)$
are on the mass shell.  The resulting wave functions satisfy,
\bea
\label{equationfn}
\left[\partial_y^2 - 2 k  \partial_y + e^{2ky} m_n^2
(1 + 2 \, r_{UV} \delta(y) + 2 \, r_{IR} \delta(y-L)) \right] f_n(y)
& = & 0 .
\eea
In fact, the introduction of opaque branes does not affect the bulk
solution for the KK modes, as written in Eq.~(\ref{eq:wavefunction}),
but instead modifies the boundary conditions at $y=0$ and $y=L$ to
reflect the discontinuity in the slope of the wave function.  Thus,
the new solutions have the same form, but with different $b$'s,
\bea
\label{boundary0}
b^0 &=&  - \frac{J_0 \left( \frac{m_{n}}{k} \right)
		 + m_{n} \, r_{UV} \, J_1 \left( \frac{m_{n}}{k} \right)}
{Y_0 \left( \frac{m_{n}}{k} \right) + m_{n} \, r_{UV} \,
Y_1 \left( \frac{m_{n}}{k} \right)}
, \\
\label{boundaryL}
b^L &=&  - \frac{J_0 \left( \frac{m_{n}}{k} e^{k L}  \right)
		 - m_{n} \, r_{IR} \, e^{kL}
J_1 \left( \frac{m_{n}}{k} e^{k L} \right)}
{Y_0 \left( \frac{m_{n}}{k} e^{k L} \right)
- m_{n} \, r_{IR} \, e^{kL} Y_1 \left( \frac{m_{n}}{k} e^{k L} \right) } ,
\label{eq:eigenmass}
\eea
indicating different admixture of the Bessel functions $J_1$ and $Y_1$
in the solutions.  Once again, we determine the quantized masses by
imposing the conditions on both boundaries, $b^0 = b^L$.  While there
is no analytic way to solve for the masses, solutions may easily be
obtained numerically.

Once the eigenmasses (and therefore $b$) have been found, we normalize
the wave functions and determine the coupling to either
brane-localized or bulk fields.  The normalization condition to
determine ${\cal N}_n$ in the presence of brane kinetic terms,
Eq.~(\ref{ortor}), may be expressed as,
\bea
\frac{1}{g_5^2} \left[r_{UV} f^2_n(0) + r_{IR} f^2_n(L) +
\int_0^L dy f^2_n(y) \right] = 1 ,
\eea
where the explicit factors of two multiplying the $r_i$ brane terms
that were introduced in Eq.~(\ref{action}) cancel as a result of
our having considered only the physical space $0 \leq y \leq L$ and
not the reflection, $y < 0$.

To examine the couplings of the KK tower to various types of fields,
either confined to branes or living in the bulk, consider some
representative interaction terms in the 5d theory,
\bea
{\cal L}_{int} & = & \int_{0}^{L} dy \left\{
\delta (y - y_\psi) \;
\left[ \overline{\psi} {\cal A}_\mu \gamma^\mu \psi \right] \right.
\nonumber \\ & &
 \hspace*{1 cm} \left.\mbox{} + \frac{1}{g_5^2}
\left( 1 + 2 \, r_{UV} \delta(y) + 2 \, r_{IR} \delta(y-L) \right)
\left[ 2 (\partial_\mu {\cal A}^a_\nu - \partial_\nu {\cal A}^a_\mu )
f^{abc} {\cal A}_b^\mu {\cal A}_c^\nu  \right]
\right. \nonumber \\
& & \hspace*{2.1cm} \left. \mbox{} + \frac{1}{g_5^2}
\left( 1 + 2 \, r_{UV} \delta(y) + 2 \, r_{IR} \delta(y-L) \right)
\left[ f^{abc} f^{ade} {\cal A}_b^\mu {\cal A}_c^\nu
{\cal A}^d_\mu {\cal A}^e_\nu \right]
\right\} ,
\eea
where we have taken fermions canonically normalized and confined to a
brane located at $y = y_\psi$.  The first term represents the gauge
boson-fermion coupling and the latter two terms are the interactions
among the bulk gauge fields for a non-Abelian theory.  In order to
derive the effective interactions between various KK modes, one
inserts the KK decomposition into this equation.

From our normalization convention for the $f_n(y)$, it follows that
the coupling of brane fields localized at $y_\psi$ to the $n$th KK
mode is
\bea
\label{eq:branecoupling}
g_n =  f_n( y_\psi ) .
\eea
Equation~(\ref{equationfn}) always has a solution with zero mass and
constant wave function.  The constant wave function insures that the
zero mode couples universally to all charged brane matter regardless
of where it is localized, as required by the unbroken gauge invariance
of the zero mode.  The normalization of the zero mode in terms of
$g_5$, $r_{IR}$, $r_{UV}$, and $L$ determines the zero mode gauge
coupling in terms of the fundamental parameters,
\bea
\label{zeromodecoupling}
g_0 &=& f_0(y) = \frac{g_5}{\sqrt{L + r_{IR} + r_{UV}}}.
\eea
Since it is the coupling of the lightest mode which we identify at low
energies with our four dimensional gauge interaction, we choose to
present the couplings of the higher KK modes relative to the zero mode
coupling.

Couplings of the higher KK modes of bulk fields are model-dependent,
being given by integrals of products of several of the wave functions.
For the three- and four-point gauge field vertices, we have,
\bea
g_{nml} & = &
\frac{1}{g_5^2}  \int_{0}^{L} dy \:
\left( 1
+ 2 \, r_{UV} \delta (y) + 2 \, r_{IR} \delta(y-L) \right) \:
\: f_n(y) \, f_m(y) \,  f_l(y) \\
g_{nmlk} & = &
\frac{1}{g_5^2}  \int_{0}^{L} dy \:
\left( 1
+ 2 \, r_{UV} \delta (y) + 2 \, r_{IR} \delta(y-L) \right) \:
f_n(y) \, f_m(y) \, f_l(y) \, f_k(y)
\eea
between the $A_n$-$A_m$-$A_l$ and $A_n$-$A_m$-$A_l$-$A_k$ modes,
respectively.  For simplicity of notation we have suppressed the
vector and color indices, but these are readily restored.  For the
vertices involving the zero mode, the results are much simpler,  since
the zero mode wave function is independent of $y$.  In the three-point
vertex, we find that the zero mode has constant coupling $g_0
\delta_{mn}$ with all other KK mode pairs.  In the four-point vertex,
setting $l=k=0$ results in a vertex factor $g^2_0 \delta_{m n}$.
Together, these results demonstrate the fact that the zero mode gauge
field's couplings take a universal form as dictated by its unbroken
gauge invariance, resulting in the same coupling to both bulk and
brane fields.  These results are the same as in the flat brane
scenario studied in \cite{Carena:2002me}.

\subsection{Opaque IR Brane}

\begin{figure}[t]
\vspace*{-1.cm}
\centerline{ \hspace*{1cm}
\epsfxsize=10.0cm\epsfysize=10.0cm
		     \epsfbox{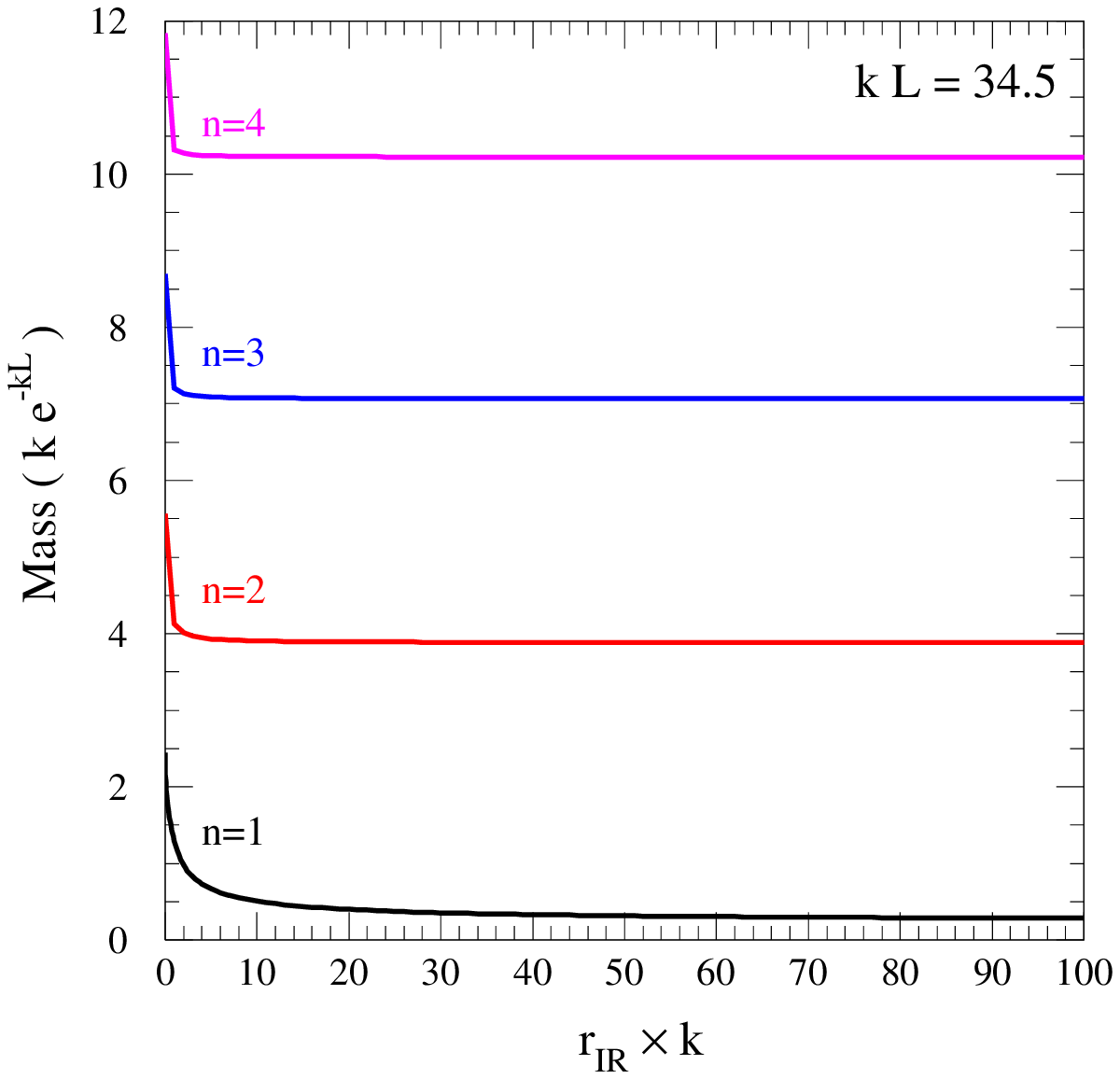} \hspace*{-1.5cm}
\epsfxsize=10.0cm\epsfysize=10.0cm
		     \epsfbox{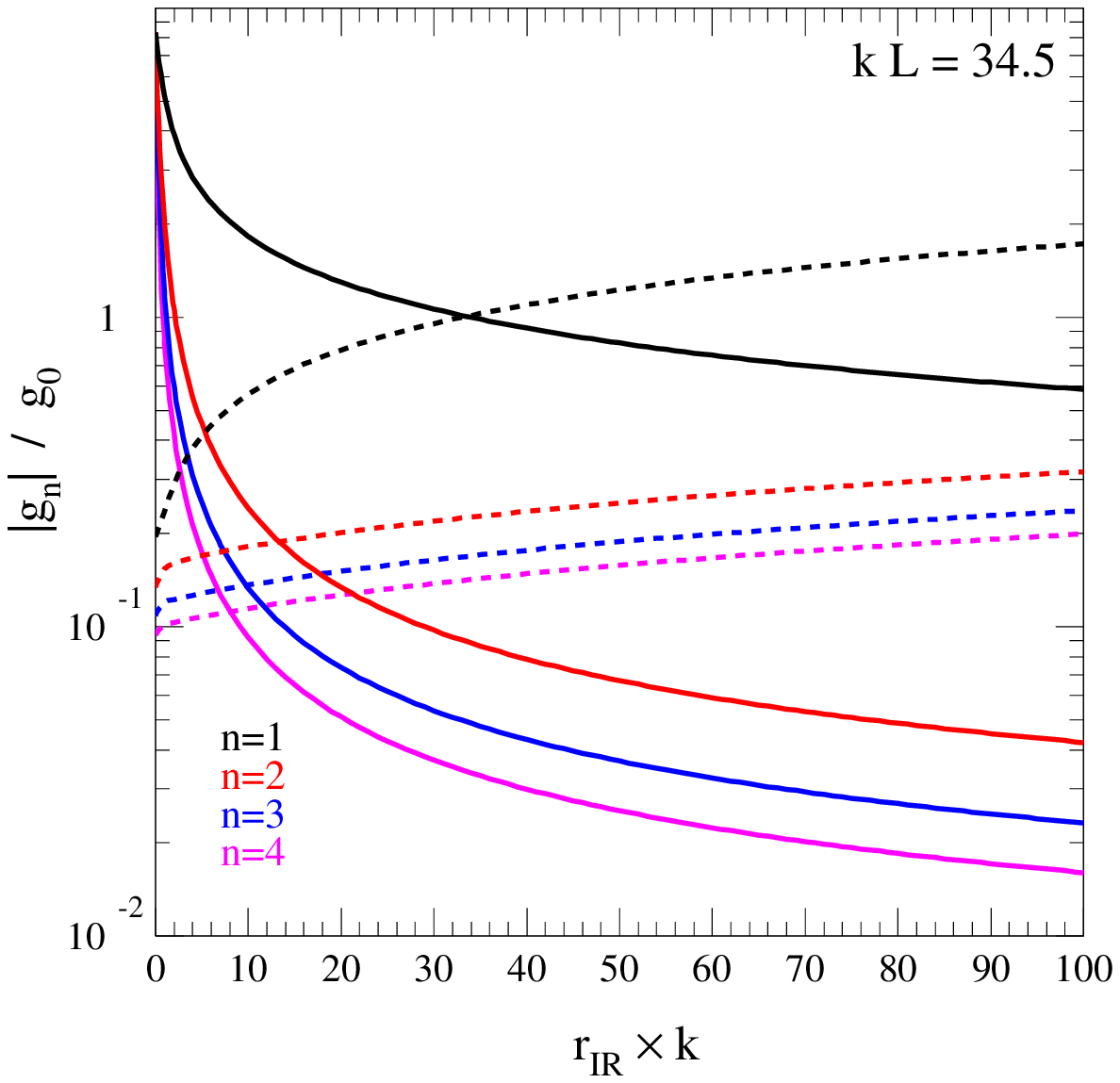}
}
\caption{The $n=1,2,3,4$ (bottom to top) KK mode masses in units of $k e^{-kL}$
and couplings to IR brane fields (solid lines) and UV brane fields
(dashed lines) relative to the zero mode coupling for the case in which the
IR brane is opaque, as a function of the opacity $r_{IR}\,k$.}
\label{fig:masses}
\end{figure}

To begin with, we consider the case where $r_{IR} \not = 0$ but
$r_{UV}=0$.  We fix $k L = 34.5$ as a typical value which generates
approximately the correct hierarchy between the Planck and weak
scales.  In Figure~\ref{fig:masses}, we present the masses of the four
lowest modes as a function of $r_{IR} \, k$.  As the brane kinetic term
increases, the KK mode masses decrease, each one approaching an
asymptotic value.  The features are qualitatively similar to the flat
space case \cite{Carena:2002me}.

Also plotted in Figure~\ref{fig:masses} are the couplings of each mode
to fields localized in either the UV or the IR branes.  In agreement
with the intuition that the localized kinetic term expels the higher
KK modes from the brane, we see that the modes decouple from the IR
brane, with the heavier modes decoupling more quickly.  (In contrast,
the couplings to the UV brane fields increase with larger $r_{IR}$.)
This behavior can also be understood from the point of view of quantum
mechanics. For sufficiently large spatial momenta, the particle loses
information about the extension of the extra dimensions and the
physics on the brane becomes determined by the value of the local
gauge coupling.  The larger the value of $r$ and the larger the
spatial momentum, the more striking this phenomenon must be.  In the
large $r$ limit, the particle propagator $G_p(L,L)$ should recover its
four dimensional behavior, and its interactions must be equivalent to
the one of a particle with gauge coupling equal to the local brane
one.  This is only possible if the higher KK modes, which dominate the
large momentum behavior of the propagator, decouple from the brane, a
behavior that is shown in Fig.~\ref{fig:masses}.

Some analytical understanding of the behavior of couplings and masses
is provided by the analysis of the five dimensional propagator for
$r_{IR} \neq 0$ and $r_{UV} = 0$.  For $p \gg k \, e^{-k L}$ the
propagator, Eq.  (\ref{veryhighEpropagator}), at $y=y'=L$ reduces to
\bea
G_p(L,L) \simeq - \frac{g_5^2 e^{kL}}{p (1 + p \, r_{IR} \, e^{kL})}.
\eea
Therefore, for values of $k \,r_{IR} \gsim 1$, one recovers a four
dimensional behavior in this energy regime,
\begin{equation}
\label{two4D}
G_p(L,L) \simeq - \frac{g_5^2}{p^2 \, r_{IR}}.
\end{equation}
This expression can be compared with the one obtained for momenta $p
\ll k \, e^{-kL}$ from Eq.~(\ref{verylowEpropagator}),
\beq
\label{one4D}
G_p(L,L) \sim - \frac{g_5^2}{p^2 (L + r_{IR})}~,
\eeq
which describes the zero mode with coupling $g_{0} = g_5/\sqrt{L +
r_{IR}}$.  The simplest interpretation of the transition from
Eq.~(\ref{two4D}) to Eq.~(\ref{one4D}) is that there is a single
massive mode with mass of order $k \, e^{-k L}$, that couples to the
brane with strength
\beq
\label{onebraneg1}
g_1 = \sqrt{\frac{L}{r_{IR}}} \; g_0~,
\eeq
while the other modes decouple.\footnote{For example, for
$k\,\LL= 34.5$, as in Fig. 1, $m_1 \rightarrow 0.24 k\, e^{-k \LL}$ as
$k r_{IR} \rightarrow \infty$.} We also see that when $r_{IR} \gg L$
then $g_{1} \ll g_{0}$, and therefore, in the large brane kinetic term
limit, only the zero mode couples to the brane with coupling
$g_{0}^{2} \approx g_{IR}^{2} = g_5^{2}/r_{IR}$.
This result agrees with the numerical results displayed in
Fig.~\ref{fig:masses}.

\subsection{Two Opaque Branes}
\label{subsec:twobranes}

The physics in the two opaque brane scenario is somewhat different
from the one brane case.  In analogy with the case of flat extra
dimensions~\cite{Carena:2002me} for values of $r_{IR}$ and $r_{UV}$
large compared to both $L$ and $1/k$, the physics at each brane must
be determined by the local couplings and, in the asymptotic limit of
$r_{IR}, r_{UV} \to \infty$ an observer on either the UV or IR brane
must be insensitive to the presence of the extra dimensions, including
the other brane.  This can only be true if, in this limit, two
massless modes appear, and appropriate linear combinations of them
couple to one of the branes with a strength given by the local
coupling, while decoupling from the other brane and vice versa.  Since
for any finite value of $r_{IR}$ and $r_{UV}$ there is only one zero
mode, what should happen is that the mass of the first KK mode tends
to zero as the local brane terms increase, while keeping a relevant
coupling to the IR and UV branes.

This behavior was already apparent from the form of the
five dimensional propagator for large values of the local brane
kinetic terms.  Consider the propagator,
$G_{p}(y,y')$ for $p \ll k e^{-k L}$, Eq. (\ref{verylowEpropagator}),
for $y=y'=L$:
\beqa
\label{gpllruv}
G_p(L,L) &\sim& - \frac{g_5^2}{p^2}
\frac{2 \, k+e^{2 k L} p^2 r_{UV}}
{2 \, k \, (L + r_{UV} + r_{IR}) + e^{2k L} \, p^2 \, r_{UV} \, r_{IR}}
\nonumber \\
&=& - \frac{g_5^2}{p^2 (L + r_{UV} + r_{IR})} -
\frac{g_5^2 \, (L + r_{UV})}
{r_{IR} \, (L + r_{UV} + r_{IR}) \,[ p^2 + m_1^2]}~.
\eeqa
The first term is the contribution of the massless zero mode, with
coupling $g_{0}$ as defined in Eq.~(\ref{zeromodecoupling}).  The
second term in Eq.~(\ref{gpllruv}) describes an additional mode of
mass $m_{1} \equiv \gamma \, k \, e^{-kL}$, where
\beq
\label{gamma}
\gamma = \sqrt\frac{2 \, (L + r_{UV} + r_{IR})}
{k \, r_{UV} \, r_{IR}}~.
\eeq
Provided that $\gamma \ll 1$, the propagator, Eq.~(\ref{gpllruv}),
shows that there is a light mode (in addition to the zero mode), whose
coupling is given by
\beq
\label{twobraneg1}
g_{1} = \sqrt{\frac{L + r_{UV}}{r_{IR}}} \, g_0~,
\eeq
which generalizes Eq.~(\ref{onebraneg1}) for nonzero $r_{UV}$.

We see that, when $r_{UV} = r_{IR} \gg L$, the light mode couples with
the same strength as the zero mode to fields localized on the IR
brane.  Notice also that, similarly to the case of vanishing $r_{UV}$,
when $r_{IR} \gg r_{UV},L$, the light mode decouples from the IR
brane.

\begin{figure}[t]
\vspace*{-1.cm}
\centerline{ \hspace*{1cm}
\epsfxsize=10.0cm\epsfysize=10.0cm
		     \epsfbox{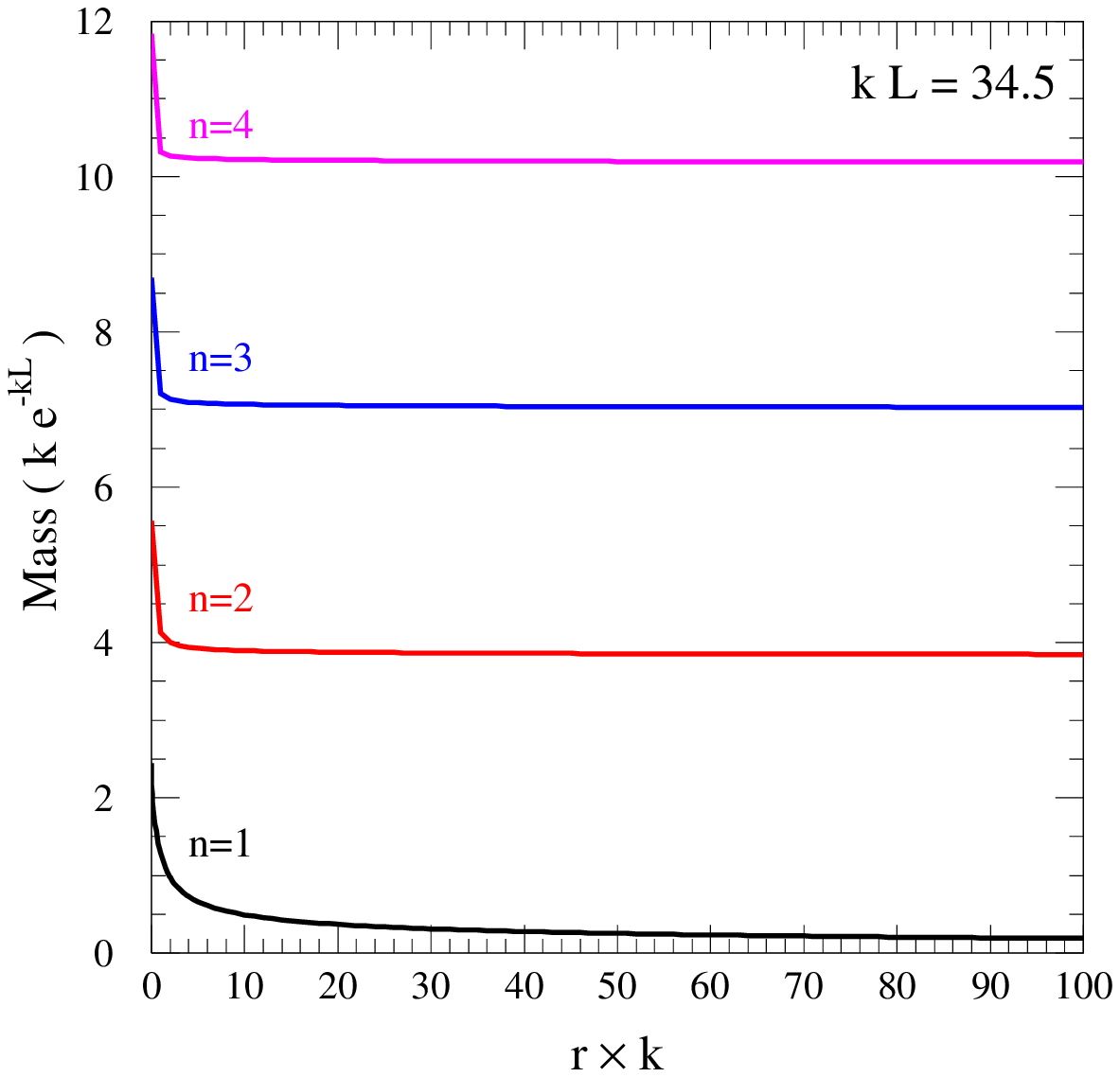} \hspace*{-1.5cm}
\epsfxsize=10.0cm\epsfysize=10.0cm
		     \epsfbox{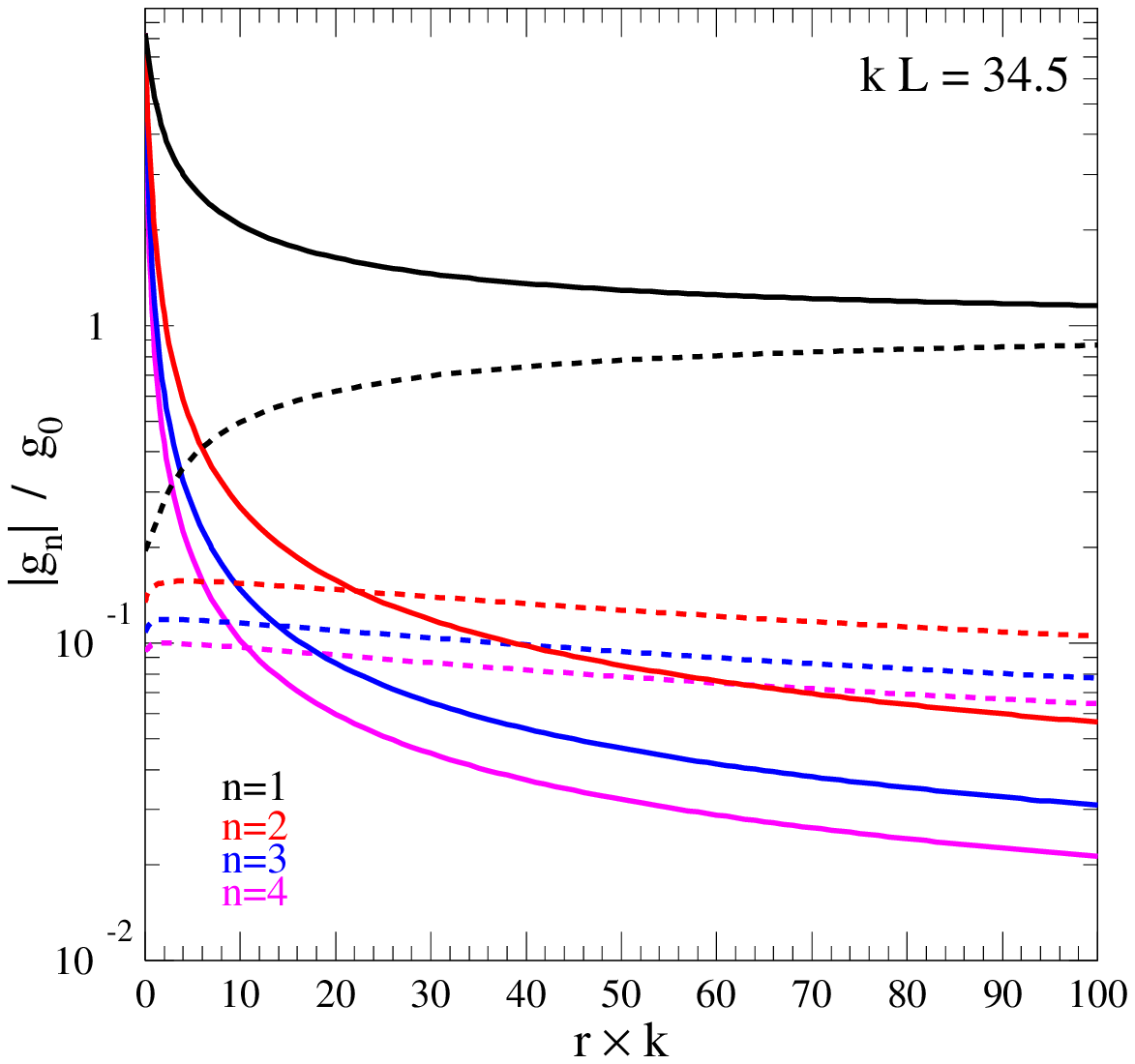}
}
\caption{The $n=1,2,3,4$ (bottom to top) KK mode masses in units of $k e^{-kL}$
and couplings to IR brane fields (solid lines) and UV brane fields
(dashed lines) relative to the zero mode coupling for the case in which both
the IR and UV branes are opaque, as a function of the common opacity 
$r\times k$.}
\label{fig:masses2}
\end{figure}

In the numerical analysis, for simplicity, we consider the case $r = 
r_{IR}= r_{UV}$.  In Figure~\ref{fig:masses2} we show the couplings to 
brane fields localized on the IR and UV branes.  The first mode mass 
goes to zero and its coupling becomes equal (and opposite in sign in 
the case of UV brane fields) to the zero mode.  This agrees with the 
limit $r \rightarrow \infty$ in which bulk propagation switches off 
and we are left with two brane gauge theories which do not interact 
with each other.  The higher modes decouple from both branes.  Note 
that the couplings to the UV brane grow slightly as $r$ increases, and 
then begin to fall off again as the higher modes decouple.

\section{Localized Higgs Effect}
\label{sec:higgs}

In order to describe the electroweak theory and address the hierarchy 
problem, the Higgs responsible for electroweak symmetry-breaking 
(EWSB) must be localized on the IR brane with action,\footnote{The
minus sign in front of the scalar kinetic term is due to our metric 
signature convention, Eq.~(\ref{lineelement}).  The factor of two in 
front of the $\delta$-function has the same origin as for the 
localized terms in Eq.~(\ref{action}).}
\bea
- \int d^4x \, dy \: \sqrt{-g} \: 2 \,\delta (y-L) \:
\left\{ (D_\mu H)^\dagger D^\mu H + \lambda \left( |H|^2 -
\frac{1}{2} v^2 \right)^2 \right\}~,
\eea
where $D^\mu = \partial^{\mu} - i A^{\mu}$ is the usual 
gauge-covariant derivative.  In the low energy effective theory (after 
rescaling the Higgs kinetic term to canonical normalization) this 
results in,
\bea
-\int d^4x \:
\left\{\eta^{\mu \nu} (D_\mu H)^\dagger D_\nu H
+ \lambda \left( |H|^2 - \frac{1}{2} v^2 e^{-2k\LL} \right)^2
\right\}~,
\eea
with the EWSB VEV red-shifted to $\vv = e^{-kL} v$.  As mentioned
above, a natural solution to the hierarchy problem is obtained for
values of $kL \simeq 34.5$.  In 5d language and working in 4d unitary
gauge, the localized VEV results in a gauge boson mass which is
itself localized on the IR brane,
\bea
\label{localmass}
- \frac{1}{2} \int d^4x \, dy \; 2\,\delta(y-L) \: \vv^2 \eta^{\mu \nu}
\A_\mu \A_\nu~.
\eea

It is therefore interesting to analyze the effects on the gauge field
propagation induced by the presence of a Higgs VEV on the IR brane.
In fact, the localized Higgs VEV bears a certain resemblance to the
local gauge kinetic terms, and it can be analyzed using similar techniques.
For the purposes of this section, we consider a simple gauge group.  We
discuss the subtleties of the $SU(2)_L \times U(1)_Y$ theory in
section~\ref{sec:pheno}.

The localized mass only affects the boundary condition to be imposed
at $y = \LL$.  Comparing Eqs.~(\ref{action}) and (\ref{localmass}) we
see that, for example, Eq.~(\ref{propboundL}) is modified to
\beq
\left[ \partial_y G_p +
e^{2\sigma} \left( r_{IR} \, p^2 + g_{5}^{2} \vv^{2} \right) G_p
\right]_{y = \LL} = 0~.
\eeq
Therefore, we can obtain $G_p(y,y')$ from Eqs.~(\ref{propagator}) and
(\ref{coefficients}) by making the replacement
\beq
\label{repl}
r_{IR} \rightarrow r_{IR} + \frac{g_{5}^{2} \vv^{2}}{p^2}~.
\eeq

The conditions for a diagonal KK decomposition may be expressed as
\bea
\label{ortorv}
\frac{1}{g_5^2} \int_{0}^{L} dy \:
f_n(y) f_m(y) \left[ 1 + 2 \, r_{UV} \delta (y) + 2 \, r_{IR} \delta (y-L) \right]
& = & \delta_{nm}, \nonumber \\
\frac{1}{g_5^2} \int_{0}^{L} dy \: e^{-2ky} \;
\left\{ f^\prime_n (y) f^\prime_m (y)
+ 2 \, g_5^2 v^2 \delta (y-L) \; f_n(y) f_m(y) \right\}
& = & m_n^2 \delta_{nm} .
\eea
Once again the KK mode wave functions can be found by requiring each
mode to satisfy the free field equation of motion for a massive vector
field.  The KK masses are determined as before, by imposing $b^0 =
b^L$, where now the coefficient $b^L$ in Eq.~(\ref{eq:wavefunction}),
determined by the boundary condition at $y = L$, can be simply
obtained from Eq.~(\ref{boundaryL}) by the replacement (\ref{repl})
with $p^{2} = - m_{n}^{2}$:
\bea
b^L &=&  - \frac{J_0 \left( \frac{m_{n}}{k} e^{k L}  \right)
- \left[ m_{n} \, r_{IR} - g_5^2 \vv^2 / m_{n} \right]\, e^{kL}
J_1 \left( \frac{m_{n}}{k} e^{k L} \right)}
{Y_0 \left( \frac{m_{n}}{k} e^{k L} \right)
- \left[  m_{n} \, r_{IR} - g_5^2 \vv^2 / m_{n} \right]
\, e^{kL} Y_1 \left( \frac{m_{n}}{k} e^{k L} \right) }~.
\eea
We will be interested in the case that $r_{UV}$ vanishes (or is small).
In this case, $b^0$ is given by
\beq
b^0 =  - \frac{J_0 \left( \frac{m_{n}}{k} \right)}
{Y_0 \left( \frac{m_{n}}{k} \right)}~.
\eeq

\subsection{Transparent IR Brane}

Let us start with the case $r_{IR} = 0$.  A first intuitive phenomenon
that occurs is that, due to energy considerations, the presence of the
VEV tries to induce a repulsion of the gauge field from the brane
location.  This is balanced by the cost in energy associated with a
non-vanishing $y$-derivative.  In a KK decomposition, this produces
mixing between the whole tower of KK modes, resulting in a deformation
of the five dimensional wave functions, including the zero mode, which
is no longer absolutely flat.\footnote{For simplicity, we continue to
refer to the lightest (would-be zero) mode as the zero mode, despite
the fact that its mass is no longer zero.}  When $v \ll k$, the cost
associated with the deformed wave function overshadows the localized
mass induced by the Higgs, and the zero mode remains approximately
flat with mass given by $g_{5} \, \vv / \sqrt{L}$.

\begin{figure}[t]
\vspace*{-1.cm}
\centerline{ \hspace*{1cm}
\epsfxsize=10.0cm\epsfysize=10.0cm
		     \epsfbox{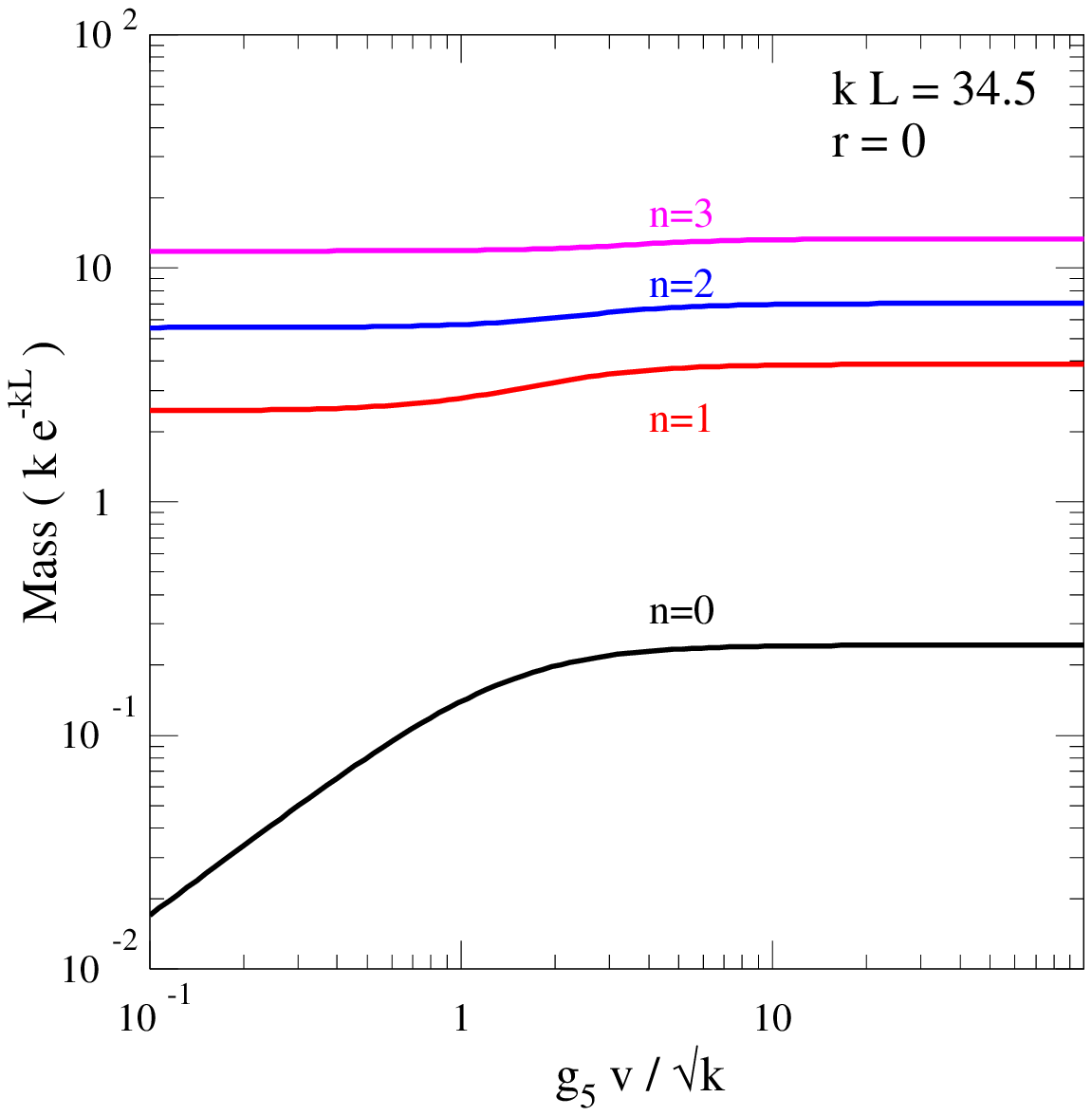} \hspace*{-1.5cm}
\epsfxsize=10.0cm\epsfysize=10.0cm
		     \epsfbox{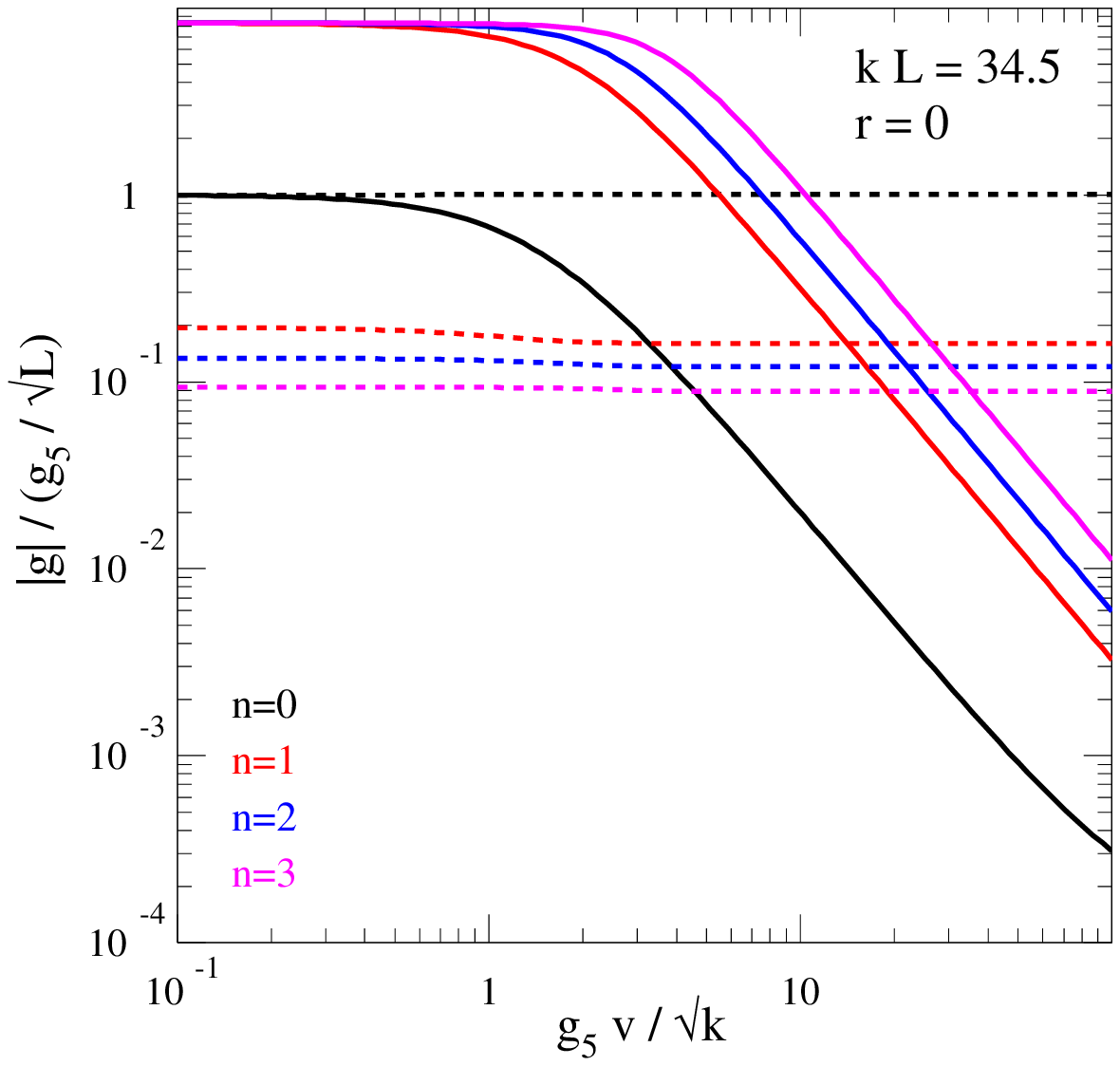}
}
\caption{The $n=0,1,2,3$ (bottom to top) KK mode masses in units of $k e^{-kL}$
and couplings to IR brane fields (solid lines) and UV brane fields
(dashed lines) relative to the bulk coupling ($g_5/\sqrt{L}$)
for the case in which a Higgs develops a VEV $v$ on the IR brane, and the
local kinetic terms are zero, as a function of the VEV, $v$}
\label{fig:massesr0v}
\end{figure}

However, when $v \gg k$, the cost associated with the derivative is 
small compared to that associated with the brane mass.  In that case 
the light KK modes are expelled from the brane.  The repulsion of the 
gauge field from the IR brane has as an immediate consequence that all 
the KK mode gauge couplings, as well as the zero mode one, tend to 
small values.  For large values of $v/k$, the ratio of the KK mode 
couplings to the zero mode tends to a constant that becomes larger 
than the value $\sqrt{2 \, k \, L}$ obtained for $v = 0$.  The second 
effect is that, due to the decoupling of the gauge field from the IR 
brane, the masses of the zero mode (and light KK modes for 
sufficiently large values of $v/k$) are no longer governed by the 
Higgs VEV. This behavior is apparent in Figure~\ref{fig:massesr0v} 
where one sees that for $v < k$ the mass of the zero mode increases 
linearly with $v$, with value $g_{5} \, \vv/\sqrt{L}$.  For $v \gsim 
k$, the mass of the lightest mode becomes more and more insensitive to 
$v$ as the mode bends away from the brane.  Also shown in 
Figure~\ref{fig:massesr0v} are the couplings of the four lowest modes 
to the IR and UV branes relative to the zero mode coupling in the 
absence of a localized Higgs VEV, $g_5 / \sqrt{L}$.  We see the 
behavior inferred above: for $v \ll k$ the coupling is close to $g_5 / 
\sqrt{L}$, while for $v \gsim k$ the KK modes are repelled from the 
brane.  This results in the higher KK modes more and more strongly 
coupled to the IR brane compared to the zero mode.  In this case, the 
theory becomes strongly coupled more quickly than it would have in the 
absence of the VEV, and the KK modes have a dramatic effect on the 
phenomenology.

\subsection{Opaque IR brane}

The situation in the case of a local gauge kinetic term on the IR
brane is somewhat different.  This is due to the fact that at
sufficiently large momenta the physics should be dominated by the
local brane term and therefore for large values of $r_{IR}$ compared
to $L$ and $1/k$ there should be a mode with coupling to the IR brane
given by the local gauge coupling and mass given approximately by this
coupling times the VEV of the Higgs field.  All other modes should
decouple from the IR brane, and their masses need not be correlated
with $v$.

We can make these observations more concrete by studying the
propagator with endpoints on the IR brane. Its limiting forms in
the high energy and low energy regimes are, respectively,
\beqa
\label{largeponebrane}
G_p(L,L) &\sim&
- \frac{g_5^2}{p \, e^{-k L} \,(1 + p \, r_{IR} \, e^{k L}) + g_{5}^{2} \vv^{2}}
\hspace{2cm} p \gg k \, e^{-k L} \\
\label{smallponebrane}
G_p(L,L) &\sim& - \frac{g_5^2}
{p^2 \left( L + r_{IR} \right) + g_{5}^{2} \vv^{2}}
\hspace{3.6cm} p \ll k \, e^{-k L}~.
\eeqa
We see that for sufficiently large momenta, $p \, r_{IR} \, e^{k L}
\gg 1$, the propagator Eq.~(\ref{largeponebrane}) reduces to $G_p(L,L)
\sim - g_{IR}^2/(p^2 + g_{IR}^2 \, \vv^2)$, where $g_{IR}^2 =
g_5^2/r_{IR}$.  Thus, when $g_{IR} \, v \gg {\rm{max}}(k,1/r_{IR})$, the
propagator describes a single
four dimensional state of mass $g_{IR} \, \vv$ (up to corrections
of order $k/(g_{IR} v$)).  It is also clear that
the energy at which this behavior sets in is lower for larger $r_{IR}
\, k$.  In fact, when $r_{IR} \gg \LL$ ($\gg 1/k$) the low energy propagator
Eq.~(\ref{smallponebrane}) takes exactly the previous form, so that
the full propagator reduces to the propagator of a single
four dimensional state with mass $g_{IR} \, \vv$ and coupling $g_{IR}$.

\begin{figure}[t]
\vspace*{-1.cm}
\centerline{ \hspace*{1cm}
\epsfxsize=10.0cm\epsfysize=10.0cm
		     \epsfbox{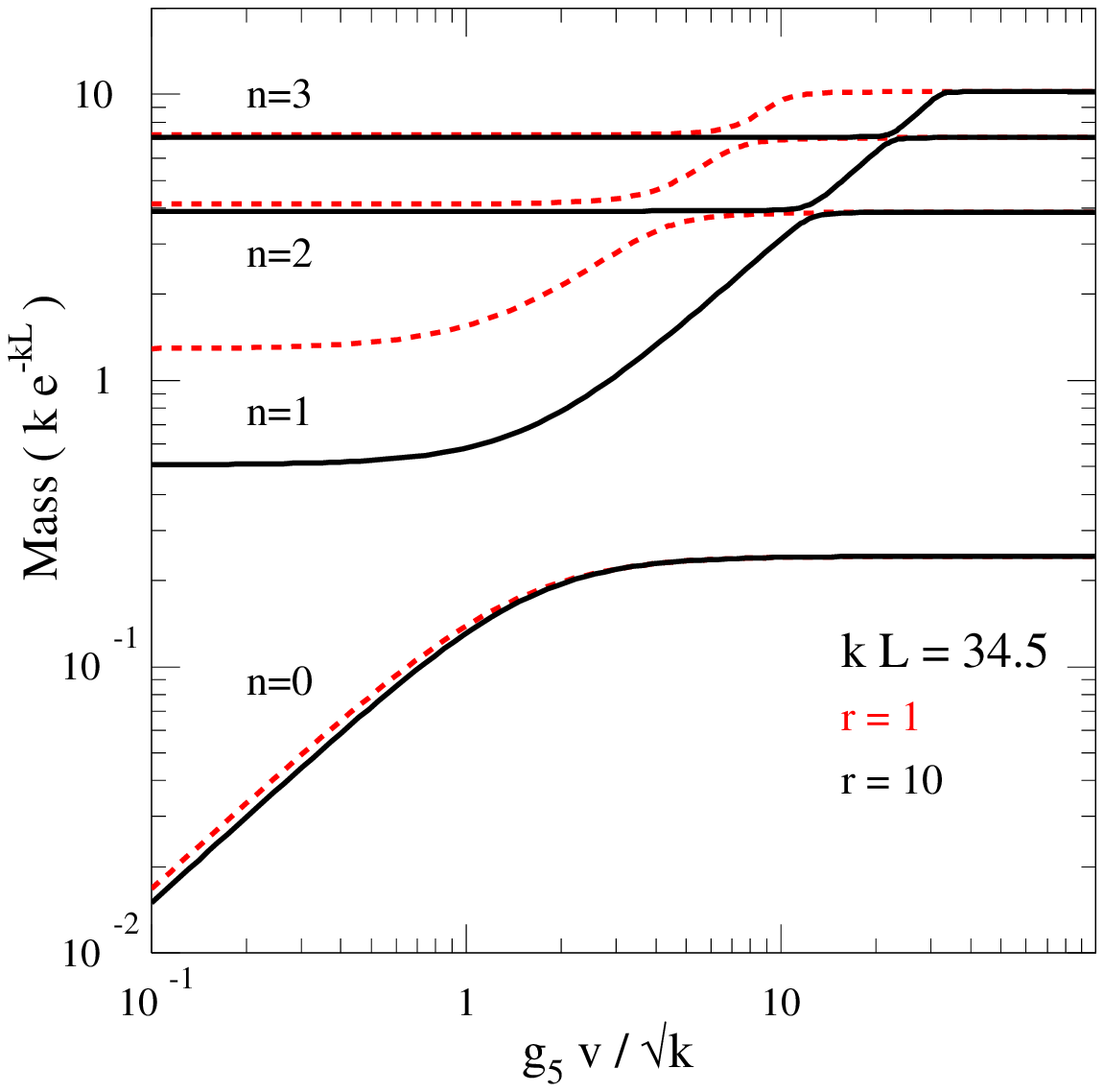} \hspace*{-1.5cm}
\epsfxsize=10.0cm\epsfysize=10.0cm
		     \epsfbox{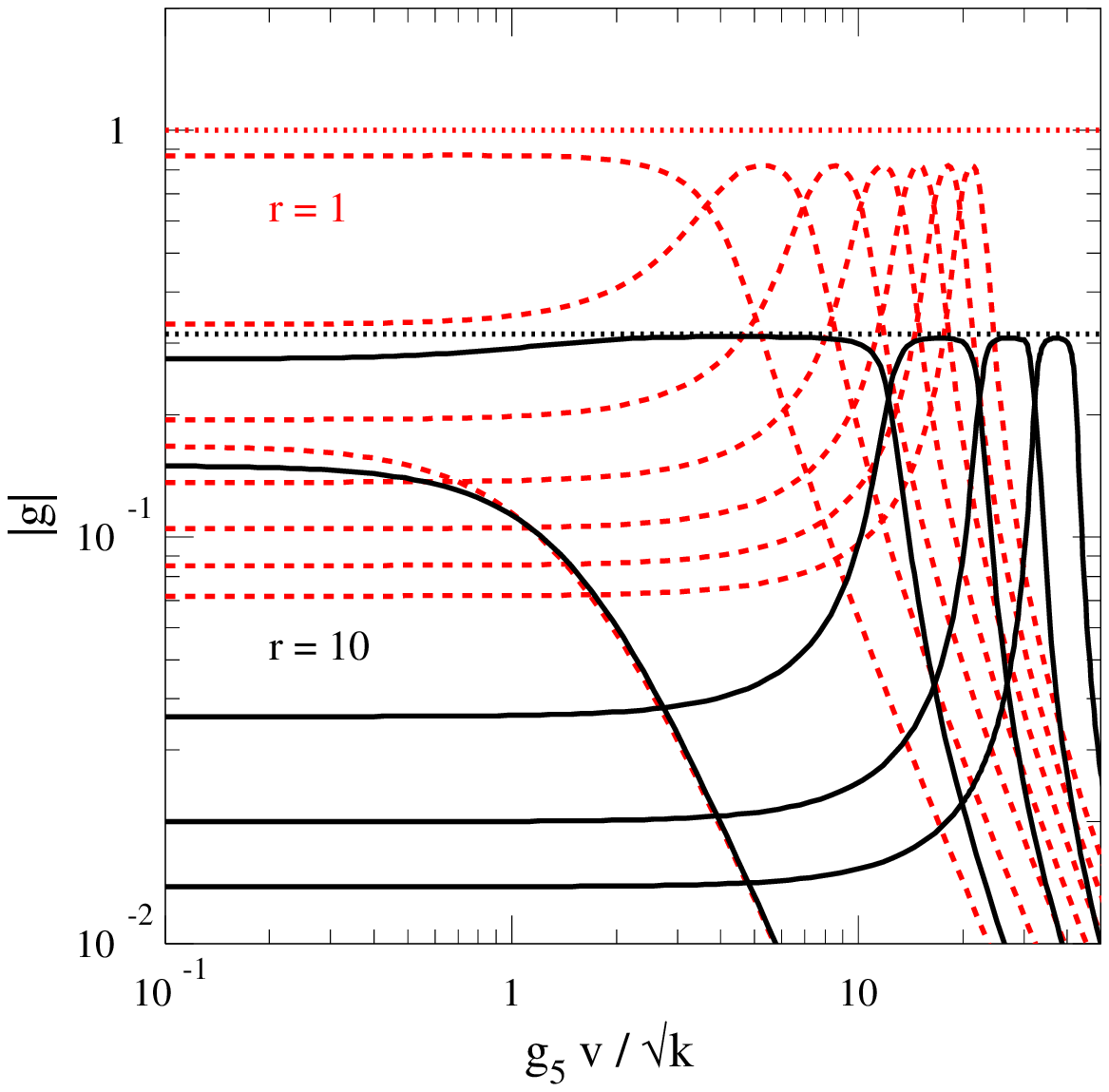}
}
\caption{The $n=0,1,2,3$ (bottom to top) KK mode masses in units of
$k e^{-kL}$ and couplings to IR brane fields (relative to $g_5 \sqrt{k}$)
for the case in which a Higgs develops a VEV on the IR brane, and the
local kinetic terms are $r_{IR} k = 1$ (red dashed curves) or $r_{IR} k = 10$
(black solid curves),
as a function of the VEV, $g_5 v /k $.  The dotted lines in the coupling
figure indicate the value of the local brane coupling (relative to
$g_5 \sqrt{k}$) corresponding to the appropriate value of $r_{IR}$.}
\label{fig:massesrv}
\end{figure}

The above described behavior is illustrated in Fig.~\ref{fig:massesrv} 
from a KK point of view, where we plot the masses and couplings (now 
relative to $g_5 \sqrt{k}$) of the four lightest modes for two 
different values of $r_{IR}$, as a function of $v/k$.  The mass 
spectrum exhibits a level-crossing phenomenon, more pronounced for 
$r_{IR} k = 10$ than for $r_{IR} k = 1$, where each mode successively 
``takes a turn'' feeling the Higgs effect.  For small $v$, it is the 
zero mode whose mass grows linearly with $v$, whereas at $v/k \sim 1$ 
the zero mode mass becomes insensitive to $v$, and it is the first KK 
mode mass which grows linearly.  This pattern repeats; as each mode 
becomes insensitive to $v$, the one above it becomes sensitive and 
grows linearly with $v$.  For large $v/k$, the transition points occur
in intervals of approximately $r_{IR} k$, with effective coupling 
approximately $g_5/\sqrt{r_{IR}}$, apart from the zero mode whose coupling
is always given by $g_5/\sqrt{L+r_{IR}}$.

For the phenomenological applications to be discussed in section 5, we
will be interested mostly in the case where $\tilde{v} \ll k\,e^{-k \LL}$.
It is clear from Eq.~(\ref{smallponebrane}) that, for $p \ll k\,e^{-k \LL}$,
there is a single state with coupling and mass given by $g_0 \simeq g_5 /
\sqrt{L+r_{IR}}$ and $m_0 \simeq g_0 \tilde{v}$, up to corrections of order
$g_5^2 v^2/k$. It will be useful to have the next order corrections
to the above parameters. We find
\beqa
\label{m0firstorder}
m_0 &=& \frac{g_5 \tilde{v}}{\sqrt{L+r_{IR}}} \left[1-\eta
\left( \frac{g_5^2 v^2}{k} \right)
+ {\cal{O}}\left( \frac{g_5^2 v^2}{k} \right)^2  \right]~, \\
\label{g0firstorder}
g_0 &=& \frac{g_5}{\sqrt{L+r_{IR}}} \left[1-2\eta \left( \frac{g_5^2 v^2}{k}
\right)
+ {\cal{O}}\left( \frac{g_5^2 v^2}{k} \right)^2 \right]~,
\eeqa
where
\beq
\label{eta}
\eta = \frac{2 k^2 \LL^2 - 2 k \LL + 1}{8 k^2 (L + r_{IR})^2}~.
\eeq
We observe that when $r_{IR} \gg \LL$ then $\eta \sim \LL^2/r_{IR}^2$, 
which shows explicitly that, provided $g_5^2 v^2/k$ is small, in the large
$r_{IR}$ limit the standard
relations among $\tilde{v}$, $g_0$ and $m_0$ are recovered.

\subsection{Two Opaque Branes}

For the sake of completeness we now consider the case where both
$r_{UV}$ and $r_{IR}$ are turned
on.  The IR brane propagator has the following limits for arbitrary
values of $r_{UV}$, $r_{IR}$ and $g_{5} \vv$:
\beqa
\label{largep}
G_p(L,L) &\sim&
- \frac{g_5^2}{p \, e^{-k L} \,(1 + p \, r_{IR} \, e^{k L}) + g_{5}^{2} \vv^{2}}
\hspace{3.8cm} p \gg k \, e^{-k L} \\
\label{smallp}
G_p(L,L) &\sim& - g_5^2
\left[ p^2 \left( r_{IR} +
\frac{2 \, k \, (L + r_{UV})}{2 \,k + e^{2 k L} \, p^2 \, r_{UV}} \right) +
g_{5}^{2} \vv^{2} \right]^{-1}
\hspace{1.1cm} p \ll k \, e^{-k L}~.
\eeqa
We see that, as required by locality, when $p \gg k \, e^{-k L}$, the 
IR propagator depends only on $r_{IR}$, not on $r_{UV}$ which is 
localized far away.  Thus, the high energy properties are the same as 
when $r_{UV} = 0$, as discussed in the previous subsection: for 
sufficiently large momenta, IR observers see a single four dimensional 
state of mass $g_{IR} \, \vv$ and coupling $g_{IR}$.

The low energy limit is more complicated and depends on the relative
size of the various localized parameters.  First, inspecting the
coefficient of the $p^{2}$ term in Eq.~(\ref{smallp}), we see that
when $p \ll k \, e^{-k L}$ there is a qualitative difference depending
on the size of $\gamma$ as defined in Eq.~(\ref{gamma}).  When $\gamma
\gg 1$, as is the case when either brane localized kinetic term
vanishes, the propagator is just
\beq
\label{zeromodepropagator}
G_p(L,L) \sim
- \frac{g_0^2}{p^2 + g_0^2 \, \vv^{2}}~,
\eeq
where $g_{0}^{2}$ was defined in Eq.~(\ref{zeromodecoupling}).
The physics is qualitatively similar to the case where $r_{UV}=0$.

The low energy physics is considerably richer when both localized 
kinetic terms are present and $\gamma \ll 1$.  In this case we have to 
be more careful about the size of the Higgs VEV relative to other 
scales.  In fact, compared to the analysis of subsection 
\ref{subsec:twobranes}, the introduction of the Higgs VEV introduces 
two new relevant scales: $g_{IR} \vv$ and $g_{0} \vv$, where 
$g_{IR}^{2} = g_{5}^{2}/r_{IR}$ and $g_{0}$ is given in 
Eq.~(\ref{zeromodecoupling}).  These two scales can be very different 
if there is a hierarchy between $r_{UV}$ and $r_{IR}$.  For 
simplicity, we will analyze here the case where $g_{0} \sim g_{IR}$.  
We find three regions according to the size of $g_{IR} v$.  When 
$g_{IR} \, v \gg k$ there are no light states in the theory and we get
an effective contact term interaction $G_p(L,L) \sim - 1/\vv^{2}$.  
When $\gamma \, k \ll g_{IR} \, v \ll k$ we find
\beq
G_p(L,L) \sim \left\{
\begin{array}{cc}
\vspace{3mm} - \frac{g_{IR}^2}{p^2}
&  \hspace{1cm} g_{IR} \, \vv \ll p \ll k \, e^{-kL}  \\
-\frac{1}{\vv^{2}}
&  \hspace{1cm} p \ll  g_{IR} \, \vv~,
\end{array}
\right.
\eeq
which shows that, below the compactification scale $k \, e^{-k L}$, there
is a single state with mass $g_{IR} \vv$, and with coupling $g_{IR}$.
Finally, when $g_{IR} \, \vv \ll \gamma \, k$, we conclude from
\beq
G_p(L,L) \sim \left\{
\begin{array}{cc}
\vspace{3mm} - \frac{g_{IR}^2}{p^2}
&  \hspace{1cm} \gamma \,  k \, e^{-kL}  \ll p \ll k \, e^{-kL}  \\
- \frac{g_0^2}{p^2 + g_0^2 \, \vv^{2}}
&  \hspace{1cm} p \ll \gamma \, k \, e^{-kL}~.
\end{array}
\right.
\eeq
that there are two light states below $k \, e^{-k L}$.  The lightest
one has mass $g_{0} \, \vv$ and coupling $g_{0}$.  The heavier one has
a mass of order $\gamma \, k \, e^{-kL}$ and coupling $g_1$ as given
in Eq.~(\ref{twobraneg1}).

\section{Electroweak Theory and Phenomenology}
\label{sec:pheno}

We now turn to the electroweak theory $SU(2)_L \times U(1)_Y$, and 
realistic phenomenology.  We assume that the Higgs and all fermions 
are confined to the IR brane, but the gauge fields are allowed to 
propagate in the bulk.  For simplicity, we treat the UV brane as 
transparent: $r_{UV} = 0$.  After EWSB, the relevant terms in the 5d
Lagrangian are thus,
\bea
{\cal L}^5_{EW} & = &
\sqrt{-g} \left\{
-\frac{1}{4 g_5^2} \W_{MN} \W^{MN} \left( 1 + 2 r_2 \delta (y - L) \right)
- \frac{1}{4 {g^\prime_5}^2} \B_{MN} \B^{MN}
\left( 1 + 2 r_1 \delta (y - L) \right)
\right. \nonumber \\ & & \hspace*{1.15cm}
\left.
\hbox{} - {v}^2 \delta (y-L) \left[
\W_M^1 \W^M_1 + \W_M^2 \W^M_2
+ \left(\W^3_M - \B_M \right) \left( \W_3^M - \B^M \right)
\right]
\right\} ~,
\eea
where the $\W_M^{(1,2,3)}$ are the three $SU(2)_L$ gauge bosons (with 
bulk coupling $g_5$), $\B_M$ is the $U(1)_Y$ gauge boson (with bulk 
coupling $g_5^\prime$), and $\W_{MN}$ and $\B_{MN}$ are the respective 
field strength tensors.  $r_1$ and $r_2$ are the IR brane gauge 
kinetic terms for the $U(1)_Y$ and $SU(2)_L$ gauge groups.

In Ref.~\cite{Csaki:2002gy}, which addressed the localized VEV effects,
but treated the brane as transparent, the situation was analyzed in the limit
$v \ll k$ (for which the symmetry-breaking effects of section~\ref{sec:higgs}
can be treated as a perturbation) by introducing the gauge rotations which
diagonalize the masses in terms of the bulk couplings,
$s \equiv g_5^\prime / \sqrt{g_5^2 + {g_5^\prime}^2}$.
Then, in the basis,
\bea
\label{eq:gaugerotation}
\W^3_\mu = c^2 \Z_\mu + \A_\mu &,& \B_\mu = -s^2 \Z_\mu + \A_\mu
\eea
the KK towers for the photon ($A$) and $Z$ decouple from each other.
For $r_1 \not = r_2$, this choice is not particularly convenient, because
although it does decouple the KK tower of the photon and $Z$ with respect
to the bulk terms, they remain mixed together by the brane terms.
Thus we consider the simplified case, $r_1 = r_2$, for which the same
rotation diagonalizes both terms.  While there is no reason why $r_1 = r_2$
should hold, and in fact since they renormalize differently, they will
certainly differ at different energy scales, it does simply illustrate the
effects of the brane kinetic terms on the phenomenological picture.

When $r_1 = r_2 = r$, the field redefinitions of Eq.~(\ref{eq:gaugerotation})
diagonalize the Lagrangian, resulting in,
\bea
\label{eq:LEW5}
{\cal L}^5_{EW} & = &
\sqrt{-g} \left\{
-\frac{s^2}{2 e_5^2} \W^+_{MN} \W_-^{MN} \left[ 1 + 2 r \delta (y - L) \right]
- \frac{1}{4 e_5^2}
\F_{MN} \F^{MN} \left[ 1 + 2 r \delta (y - L) \right]
\right. \nonumber \\ & & \hspace*{-.3cm}
\left.
\hbox{} -\frac{s^2 c^2}{4 e_5^2}
\Z_{MN} \Z^{MN} \left[ 1 + 2 r \delta (y - L) \right]
- 2 {v}^2 \delta (y-L) \left( \W^+_M \W_-^M
+ \frac{1}{2} \Z_M \Z^M \right)
\right\} ,
\eea
where we have introduced $1 / e_5^2 = 1 / g_5^2 + 1 / {g_5^\prime}^2$, the 5d
photon coupling.

We can now make contact with the results of section~\ref{sec:higgs}.  
Clearly the zero mode of the photon is associated with the gauge 
interactions we see at low energies.  The weak gauge bosons are 
somewhat more subtle.  In section~\ref{sec:higgs} we saw that for $r 
\ll 1/k$ the first few KK modes couple a factor of order 10 (or more 
for large $v/k$) more strongly to the IR brane than the zero mode, and 
have approximately equal spacing.  This suggests that for a viable 
phenomenological picture, it is the zero modes that should be 
associated with the weak gauge bosons observed in experiments, with 
the higher KK modes suitably heavy such that they evade current 
experimental limits.

However, for large $r$, there is only ever one gauge boson with 
relevant gauge coupling to the fermions.  Provided the lighter modes 
are sufficiently weakly coupled, they could have escaped detection up 
until now.  Thus, for $r \sim 1/k$ we have a choice as to which mode 
plays the role of the observed weak bosons, at least for some range of 
parameters.  For simplicity, we will restrict our attention to the 
simpler case in which we associate the known weak interactions with 
the zero modes of the $\W$ and $\Z$ fields, and leave the more exotic 
case in which we may actually be identifying the higher KK modes as 
the mediators of the weak force for future work.

\subsection{Matching to the Effective Theory}

In the Standard Model the physics predominantly depends on three 
parameters of the electroweak theory, $e$, $\sin^2 \theta_W$, and $G_\mu$.  
The accuracy of the data is such that loop-level effects are 
important, introducing a strong dependence on the top mass ($m_t$), 
and relevant dependence on the strong coupling at the $Z$ pole 
($\alpha_S$) and Higgs mass ($m_h$).  The usual procedure is to use
the three most precisely measured quantities, the Fermi constant from
muon decay ($G_\mu$), the mass of the $Z$ boson ($M_Z$), and the 
electromagnetic coupling at the $Z$-pole ($\alpha_Z$), to fix the 
three tree-level parameters, and then to combine the sum of the 
precision data in a fit to either $m_h$ (and $m_t$ and $\alpha_S$),
or, allowing for nonstandard corrections to the weak boson 
self-energies (oblique corrections), to fit the oblique parameters 
$S$, $T$, and $U$ \cite{Peskin:1991sw}.

This procedure is not completely appropriate whenever there are 
important non-oblique corrections which directly modify the fermion 
couplings, or which modify the self-interactions of the gauge bosons 
beyond what is contained in the oblique parameters.  However, it is 
still sensible whenever the oblique corrections capture the dominant 
new physics contributions.  In the specific case of the RS model with 
bulk gauge bosons, one immediately encounters a problem related to 
non-oblique corrections.  The muon decay now proceeds through the 
entire tower of KK modes of the $W$.  This was handled in Refs.  
\cite{Davoudiasl:1999tf,Csaki:2002gy,Rizzo:1999br}~\footnote{See also 
Ref.~\cite{Huber:2000fh}, and Ref.~\cite{Burdman:2002gr} for a 
description of bulk fermion effects.} by introducing an additional fit 
parameter $V$ which measures the amount of contribution to $G_\mu$ 
from the higher KK modes relative to the zero mode.  However, as we 
will see below, we will find it more convenient to avoid introducing 
$V$, and instead define effective parameters, $S$, 
$T$, and $U$, which in practice take into account all relevant
oblique and non-oblique corrections necessary to describe the Z-pole 
precision electroweak observables and the W mass.

At tree level, our model point is specified in terms of the Lagrangian 
in Eq.~(\ref{eq:LEW5}) by the six quantities $e_5$, $s$, $v$, $r$, $L$, 
and $k$.  $k$ can be taken to be the 5d Planck scale, and thus may be
thought of as setting the overall dimensionful scale.  We must 
identify the 4d quantities $e$, $\sin^2 \theta_W$, and effective 4d
Higgs VEV $\tilde{v}$ from the measurements of $\alpha_Z$, $M_Z$, and $G_\mu$.
This will determine three of our six parameters in terms of the other 
three.  We represent this freedom by treating $r$ and $L$ as the free 
parameters, and use the input data to specify $e_5$, $s$, and $v$ in 
terms of them.

Applying the results of section~\ref{sec:higgs}, we derive the effective
Lagrangian for the zero modes,
\bea
-\frac{1}{2} W^+_{\mu \nu} W_-^{\mu \nu}
-\frac{1}{4} Z_{\mu \nu} Z^{\mu \nu}
- \frac{1}{4} F_{\mu \nu} F^{\mu \nu}
- m_W^2 W^+_\mu W_-^\mu
- \frac{m_Z^2}{2} Z_\mu Z^\mu \\
\mbox{}+\frac{1}{\sqrt{2}} f_{W}
\left( \overline{\psi} \gamma^\mu T_{+} \psi
W_{\mu}^{+} + \overline{\psi} \gamma^\mu T_{-} \psi W_{\mu}^{-}
\right) + f_{Z} \overline{\psi} \gamma^\mu (T_{3} - s^{2} Q) \psi
Z_\mu + f_{A} \overline{\psi} \gamma^\mu Q \psi A_\mu~,
\eea
where $f_{W}$, $f_{Z}$ and $f_{A}$ are the $W$, $Z$ and
photon zero mode wave functions evaluated at the IR brane, $y=L$,
$\psi$ stands for the brane-localized fermions, $T_\pm$, $T_3$ are
the relevant weak isospin matrices and $Q$ is the electric charge.  Note
also that here
and below, 4d indices are raised and lowered by the canonically
normalized Lorentz metric, $\eta_{\mu \nu}$.  The quantities $m_W$ and
$m_Z$ are determined numerically for a given choice of $v$, $e_5$, $s$,
$r$, $L$, and $k$, as in Fig.~\ref{fig:massesrv}.  Note that this
implies that these quantities have no direct relation to those which
one would have expected in the SM.

Since the photon experiences no symmetry breaking, the zero mode wave function
is flat, and thus we can identify the 4d electromagnetic coupling $e$ with,
\beq
e \equiv f_{A} = \frac{e_5}{\sqrt{L + r}} ~.
\eeq
This allows us to reproduce the electromagnetic coupling 
$\alpha_Z^{-1} = 128.92(3)$.  The muon decay constant $G_\mu$ should 
be identified with the (full) $W$ boson at zero momentum transfer.  
Thus, it implicitly contains the effects of the entire KK tower.  From 
Eq.~(\ref{smallponebrane}) we see that at zero momentum transfer, the 
tree-level gauge boson propagator reduces to the very simple form,
\bea
\label{GFermi}
- G(p^2 = 0;L,L) & = & \frac{f_{W}^{2}}{m_{W}^{2}} +
\sum_{n \neq 0}\frac{f_{W_{n}}^{2}}{m_{W_{n}}^{2}} ~ = ~ 
\frac{1}{\tilde{v}^2} ~ = ~ 4 \sqrt{2} G_{\mu} ~,
\eea
which implies that fixing $G_\mu$ to its experimentally measured value
of $G_\mu = 1.16639(1) \times 10^{-5}$ Ge${\rm V}^{-2}$ determines 
$\vv\simeq 123$ GeV. This result for the propagator is exact at $p^2 = 0$
and includes the sum of the contributions of all the KK modes of the
charged weak gauge bosons.

The final quantity we would like to use to determine the input
parameters is the $Z$ mass.  Having already determined $e_5$ and $v$
(as functions of $r$ and $L$), we can accomplish this by adjusting $s$
to the value which produces the correct $m_Z$.  We find this value
numerically as in section~\ref{sec:higgs}, adjusting the quantity 
$v e_5 / s c$ until the mass of the lowest mode is the $Z$ boson mass
measured in experiments, $M_Z = 91.1875(21)$.  The advantage of
matching directly to $\alpha_Z$, $M_Z$, and $G_\mu$ is that, with
appropriate definitions, we can carry over the established machinery
for electroweak fits, and use the usual bounds on $S$, $T$, and $U$ to
directly constrain the RS model, without refitting the data (see
below).

It remains to determine the effect on precision observables as a
function of $r$ and $L$, which allows us to determine the region of $r$
and $L$ consistent with experimental data.  Since the gauge boson
interactions with the fermions are affected in a universal way, most of
the effects may be captured by effective $S$, $T$ and $U$ parameters.
The primary remaining non-oblique effect is the coupling strengths of
the self-interactions of the zero mode $W$'s and $Z$.  These depend on
integrals over $y$ of three or four gauge boson wave functions.  Thus,
while they are determined from the $f(y)$, they are not related to them
in a simple way.  However, as these interactions are only mildly
constrained by LEP-2 data \cite{LEPTCG}, the resulting constraints are
very weak.

We will find it convenient to rescale the zero mode wave functions,
\bea
f_A & = & \frac{e_5}{\sqrt{L+r}} \hat{f}_A \\
f_Z & = & \frac{e_5}{s\,c \sqrt{L+r}} \hat{f}_Z
= \sqrt{g^2 + {g^\prime}^2} \hat{f}_Z  \\
f_W & = & \frac{e_5}{s \sqrt{L+r}} \hat{f}_W
= g  \hat{f}_W
\eea
for which the deviation from the tree-level SM predictions appears as a
deviation of $\hat{f}$ from unity.  Note that given our matching
prescription we have $\hat{f}_A = 1$.  We have defined $g$ and
$g^\prime$ with respect to $s$ and $c$, the sine and cosine of the
gauge rotation angle defined above Eq.~(\ref{eq:gaugerotation}).  This
angle is different from the weak mixing angle implicit from our choice
of input parameters,
\bea
s_0^2 \; c_0^2 & = & \frac{\pi \alpha_Z}{\sqrt{2} G_\mu m_Z^2}.
\eea

In order to make contact with the oblique parameters, we further 
rescale the fermion interactions to unity, resulting in the 
self-energy part of the effective Lagrangian taking the form,
\bea
& &
-\frac{1}{2 g^2 \hat{f}^2_W} W^+_{\mu \nu} W_-^{\mu \nu}
-\frac{1}{4 (g^2 + {g^\prime}^2) \hat{f}^2_Z} Z_{\mu \nu} Z^{\mu \nu}
- \frac{1}{4 e^2} F_{\mu \nu} F^{\mu \nu} \nonumber \\ & &
- \frac{m_W^2}{g^2 \hat{f}^2_W} W^+_\mu W_-^\mu
- \frac{m_Z^2}{2 (g^2+{g^\prime}^2) \hat{f}^2_Z} Z_\mu Z^\mu ~.
\label{eq:weakzeromodes}
\eea

\subsection{The Effective $S$, $T$, and $U$ Parameters}

The tree-level contributions to $S$, $T$, and $U$ can be simply 
determined by matching the effective Lagrangian for the zero modes, 
Eq.~(\ref{eq:weakzeromodes}) to a generic effective Lagrangian 
including a parameterization of the oblique corrections,
\bea
\label{eq:Leweff}
{\cal L} & = &
-\frac{1}{2 g^2} \left( 1 - \Pi^\prime_{WW} \right)
W^+_{\mu \nu} W_-^{\mu \nu}
-\frac{1}{4 (g^2 + {g^\prime}^2) }
\left( 1 - \Pi^\prime_{ZZ} \right)
Z_{\mu \nu} Z^{\mu \nu}
\nonumber \\ & &
\mbox{} - \frac{1}{4 e^2} \left( 1 - \Pi^\prime_{\gamma \gamma} \right)
F_{\mu \nu} F^{\mu \nu}
- \frac{s\,c}{2 e^2} \Pi^\prime_{\gamma Z} F^{\mu \nu} Z_{\mu \nu}
- \left( \widetilde{v}^2 + \frac{1}{g^2} \Pi_{WW}(0) \right) W^+_\mu W_-^\mu
\nonumber \\ & &
\mbox{} - \frac{1}{2}
\left( \widetilde{v}^2 + \frac{1}{(g^2 + {g^\prime}^2)} \Pi_{ZZ}(0) \right)
Z_\mu Z^\mu \nonumber \\ & &
\mbox{} + \frac{1}{\sqrt{2}} \left( \overline{\psi} \gamma^\mu T_{+} \psi
W_{\mu}^{+} + \overline{\psi} \gamma^\mu T_{-} \psi W_{\mu}^{-}
\right) + \overline{\psi} \gamma^\mu (T_{3} - s^{2} Q) \psi
Z_\mu + \overline{\psi} \gamma^\mu Q \psi A_\mu~,
\eea
where $\widetilde{v}$ plays the role of the Higgs VEV extracted from
$G_\mu$, Eq~.~(\ref{GFermi}). We have followed Ref.~\cite{Csaki:2002gy}
and written these in standard self-energy notation, despite the fact
that the 5d contributions are in fact tree level.  The ADS/CFT 
correspondance suggests that the 5d theory is dual to a 4d theory of
walking Technicolor \cite{Arkani-Hamed:2000ds}.  Thus, the tree level
5d contributions correspond to the loop-level corrections in the dual 
theory \cite{Csaki:2002gy}.  Matching Eqs.~(\ref{eq:weakzeromodes}) 
and (\ref{eq:Leweff}), we obtain,
\bea
\frac{\overline{\Pi}^\prime_{WW}}{g^2} & = & \overline{\Pi}^\prime_{11} =
\frac{1}{g^2} \left( 1 - \frac{1}{\hat{f}_W^2} \right) \\
\frac{\overline{\Pi}^\prime_{ZZ}}{\left( g^2 + {g^\prime}^2 \right)} & = &
\overline{\Pi}^\prime_{33} = \frac{1}{\left( g^2 + {g^\prime}^2 \right)}
\left( 1 - \frac{1}{\hat{f}_Z^2} \right) \\
\overline{\Pi}^\prime_{3Q} & = & \overline{\Pi}^\prime_{QQ} = 0 \\
\frac{\overline{\Pi}_{WW}(0)}{g^2} & = & \overline{\Pi}_{11}(0) =
\frac{m_W^2}{g^2 \hat{f}_W^2}
- \widetilde{v}^2 \\
\frac{\overline{\Pi}_{ZZ}(0)}{(g^2 +  {g^\prime}^2)} & = &
\overline{\Pi}_{33}(0) =
\frac{m_Z^2}{(g^2 +  {g^\prime}^2) \hat{f}_Z^2}
- \widetilde{v}^2 ~,
\eea
where the $\overline{\Pi}$ refer to the fact that these are only the
tree-level self-energy contributions from the extra-dimensional effects.
The full $\Pi$ will also include both loop-level SM contributions,
and also loop-level extra dimensional ones.

The usual oblique parameters $S$, $T$, and $U$ are defined in terms of the
self-energies by,
\bea
S & \equiv & 16 \pi \left( \Pi^\prime_{33} - \Pi^\prime_{3Q} \right) \\
T & \equiv & \frac{4 \pi}{s^2 c^2 M_Z^2}
\left( \Pi_{11}(0) - \Pi_{33}(0) \right) \\
U & \equiv & 16 \pi \left( \Pi^\prime_{11} - \Pi^\prime_{33} \right) ~,
\eea
where $M_Z$ is the $Z$ mass from the SM relations in terms of $e$,
$s_0$ and $\tilde{v}$.  Given our matching conditions, $m_Z = M_Z$.
Thus, we derive the tree level 5d contributions to the usual $S$, $T$,
and $U$,
\bea
\label{eqS}
\overline{S} & = & \frac{4 s^2 c^2}{\alpha}
\left( 1 - \frac{1}{\hat{f}_Z^2} \right) \\
\label{eqT}
\overline{T} & = & \frac{1}{\alpha}
\left( \frac{m_W^2}{c^2 m_Z^2 \hat{f}_W^2}
- \frac{1}{\hat{f}_Z^2} \right) \\
\label{eqU}
\overline{U} & = & \frac{4 s^2}{\alpha} \left[
1 - \frac{1}{\hat{f}_W^2} - c^2
\left( 1 - \frac{1}{\hat{f}_Z^2} \right) \right]
\eea
where once again, the barred quantities indicate that these are only the
contributions from the tree level extra-dimensional effects.

At the end of subsection 4.2, we found approximate expressions for the 
mass and coupling to IR fields of the zero mode gauge bosons, that are 
valid when $\epsilon \equiv g_{5}^{2} v^{2}/k \ll 1$ (see 
Eqs.~(\ref{m0firstorder}) and (\ref{g0firstorder})), where $g_5$ 
stands for the appropriate gauge coupling ($g_5$ for the charged $W$ 
bosons and $\sqrt{g_5^2 + {g'}_5^2}$ for the $Z$).  It will be useful to
have analytical expression for $S$, $T$ and $U$ to the same order.  In 
our approach we determine the sine of the rotation angle between the 
gauge and mass eigenbasis, $s$, by requiring that the $Z$ mass is 
reproduced.  This angle is in general different from the standard 
model value, but when $\epsilon \ll 1$ we expect them to differ by 
order $\epsilon$.  Using Eq.~(\ref{m0firstorder}) with the general 
$g_{5}$ of section 4 replaced by $e_{5}/(s\,c)$ we have
\beqa
m_{Z} &=& \frac{e \tilde{v}}{s\,c}
\left( 1 - \eta\,\epsilon + \cdots \right)
\equiv \frac{e \tilde{v}}{s_{0}\,c_{0}}~, \nonumber
\eeqa
where $\eta$ was defined in Eq.~(\ref{eta}) and
$\epsilon = e_{5}^{2} v^{2}/(s^{2} c^{2} k)
\approx e_{5}^{2} v^{2}/(s_{0}^{2} c_{0}^{2} k)$.
The above relation gives
\beq
\label{deltasine}
s = s_{0} \left(1 - \frac{c_{0}^{2}}{c_{0}^{2}-s_{0}^{2}} \eta \,
\epsilon + \cdots \right)~,
\eeq
where
\beq
\eta\,\epsilon = \frac{2 k^2 \LL^2 - 2 k \LL + 1}{8 k (L + r_{IR})}
\frac{e^{2} v^{2}}{s_{0}^{2} c_{0}^{2} k^2}~.
\eeq
We can also write $M_{W}$, $\hat{f}_{Z}(\LL)$ and $\hat{f}_{W}(\LL)$ as:
\beqa
\label{mWapprox}
m_{W} &=& \frac{e\tilde{v}}{s} \left[1 - c^{2} \eta\,\epsilon
+ {\cal{O}}(\epsilon^{2}) \right] =  \frac{e
\tilde{v}}{s_{0}} \left[1 + \frac{2s_{0}^{2}c_{0}^{2}}{c_{0}^{2}-s_{0}^{2}}
\eta\,\epsilon + {\cal{O}}(\epsilon^{2})\right]~, \\
\label{fZapprox}
\hat{f}_{Z}(\LL) &=& 1 - 2 \eta\,\epsilon + {\cal{O}}(\epsilon^{2})~, \\
\label{fWapprox}
\hat{f}_{W}(\LL) &=& 1 - 2 c_0^2 \eta\,\epsilon + {\cal{O}}(\epsilon^{2})~.
\eeqa
Plugging these expressions in Eqs.~(\ref{eqS})-(\ref{eqU}), we find
\beqa
\overline{S} &=& -\frac{16 s_{0}^{2}c_{0}^{2}}{\alpha} \eta\,\epsilon
+ {\cal{O}}(\epsilon^{2})~, \\
\overline{T} &=& -\frac{2 s_{0}^{2}}{\alpha} \eta\,\epsilon +
{\cal{O}}(\epsilon^{2})~, \\
\overline{U} &=& {\cal{O}}(\epsilon^{2})~.
\eeqa

Due to the presence of non-oblique corrections associated with heavy 
KK modes, the $S$, $T$ and $U$ parameters defined above are not
sufficient to describe the precision electroweak observables.
However, as observed in Ref.~\cite{Csaki:2002gy}, the Z-pole
precision observables may be properly described by the introduction
of an effective  parameter, $T_{\rm{eff}}$, which includes the non-oblique
corrections to the weak mixing angle,
\beq
\label{newdeltasine}
s^{2}-s_{0}^{2} = \frac{\alpha}{c^{2}-s^{2}} \left( \frac{1}{4} S -
s^{2} c^{2} T_{\rm{eff}} \right) ~,
\eeq
where $T_{\rm{eff}}$ is given by
\beq
T_{\rm{eff}} = T + \Delta T
\eeq
and
\beq
\Delta T = -\frac{1}{\alpha} \frac{\delta G_\mu}{G_\mu} =
-\frac{1}{\alpha}
\sum_{n \neq 0} \left(\frac{f_{W_{n}}^{2}}{f_{W}^{2}} \right)
\left( \frac{m_{W}^{2}}{m_{W_{n}}^{2}} \right) = -\frac{1}{\alpha}
\left( \frac{m_{W}^{2}}{\tilde{v}^{2} f_{W}^{2}} - 1 \right) =
- \frac{2 c_0^2}{\alpha} \eta\,\epsilon + {\cal{O}}(\epsilon^{2})~,
\eeq
where $\delta G_\mu$ is the non-oblique contribution to the muon decay constant due
to the exchange of KK modes.
With this definition of $T_{\rm{eff}}$, we can recover
Eq.~(\ref{deltasine}), using Eqs.~(\ref{mWapprox}) and
(\ref{fWapprox}).  Note that in the RS theory, the bare parameter $s$
coincides with the Kennedy-Lynn running coupling $s_{*}$ \cite{Kennedy:1988sn}.

The non-oblique corrections to $G_\mu$ affect also the expression of $m_W/m_Z$,
but the relative coefficient between the oblique and non-oblique corrections
encoded in $T_{\rm{eff}}$ is different from the one appearing in $s^2$.
Therefore, to properly parameterize $m_{W}^{2}/m_{Z}^{2}$ in terms
of effective parameters $S_{\rm{eff}} = S$, $T_{\rm{eff}}$ and $U_{\rm{eff}}$,
\beq
\label{newmW}
\frac{m_{W}^{2}}{m_{Z}^{2}} - c_{0}^{2} = \frac{\alpha c^{2}}{c^{2}-s^{2}} 
\left( - \frac{1}{2} S_{\rm{eff}} + c^{2} T_{\rm{eff}} +
\frac{c^{2}-s^{2}}{4s^{2}} U_{\rm{eff}} \right) ~,
\eeq
one can introduce 
\beq
U_{\rm{eff}} = U - 4 s^{2} \Delta T~.
\eeq
Observe that the above expressions, Eqs.~(\ref{newdeltasine}) and
(\ref{newmW}) reproduce Eqs.~(\ref{deltasine}) and
(\ref{mWapprox}), respectively.\footnote{The parameterization for
$m_W^2/m_Z^2$ in terms of the effective parameters $S_{\rm{eff}}$,
$T_{\rm{eff}}$ and $U_{\rm{eff}}$ differs from the one presented in the
appendix of Ref.~\cite{Csaki:2002gy} by the additional contribution
from $U_{\rm{eff}}$.}

The above parameterization serves to describe all Z-pole observables as
well as the $W$ mass.  Following Ref.~\cite{Altarelli:2001wx}, we shall
use a fit to these observables to place limits upon the free parameters
of our theory.

In terms of the expansion in $\epsilon$, we find that the tree-level
five-dimensional contributions, including the non-oblique corrections
to $G_\mu$, can be given by
\bea
\label{Seff}
\overline{S}_{\rm{eff}} &=& -\frac{16 s_{0}^{2}c_{0}^{2}}{\alpha} \eta\,\epsilon
+ \ldots  \approx - 366\,\eta\,\epsilon~, \\
\label{Teff}
\overline{T}_{\rm{eff}} &=& - \frac{2}{\alpha} \eta\,\epsilon + \ldots
\approx - 258\,\eta\,\epsilon~,\\
\label{Ueff}
\overline{U}_{\rm{eff}} &=& \frac{8 s_0^2 c_0^2}{\alpha} \eta\,\epsilon
+ \ldots  \approx 183\,\eta\,\epsilon~,
\eea
We see that $\overline{S}_{\rm{eff}}$ and $\overline{T}_{\rm{eff}}$ are
negative while $\overline{U}_{\rm{eff}}$ is positive (and not small).

Before closing this section, let us stress that, following
Ref.~\cite{Csaki:2002gy}, we are ignoring loop-level contributions to
the precision electroweak observables, including those induced by a
potentially light radion.  The computations of Ref.~\cite{Csaki:2000zn}
indicate that, unless there is strong mixing between the radion and the
Higgs field~\cite{Giudice:2000av}, these corrections are small and thus
may be typically neglected.  We have also checked that for the
parameters that give a good description to the data, the effect of the
tree-level neutral KK mode interactions in the Z-pole observables is
unobservably small.

\subsection{Comparison with Data}

The SM with a light Higgs, with mass of about 120 GeV, provides an 
excellent description of the precision electroweak observables 
measured at the LEP, Tevatron and SLD colliders.  While 
$\sin^2\theta_{eff}$ plays a key role in the determination of the 
above quoted Higgs mass range, there is a discrepancy of more than about 
3$\sigma$'s between the value of the weak mixing angle extracted from 
the lepton asymmetries and the one extracted from the hadron 
asymmetries.  A good fit to the hadron asymmetries, which are 
dominated by the forward-backward asymmetry of the $b$-quark, 
$A_{FB}^b$, tends to require much larger values of the Higgs mass.  On 
the other hand, a good fit to the lepton asymmetries leads to the 
preference of a Higgs mass value of about 50 
GeV~\cite{Chanowitz:2001bv}, well below the present direct search 
limit.  In fact, the internal consistency of the precision electroweak 
data is dramatically improved by the exclusion of the hadron 
asymmetries: While the fit to the combined data from these colliders 
has a confidence of about $4\%$ 
\cite{Chanowitz:2001bv,Altarelli:2001wx},
it increases to more than 30$\%$ once the hadronic asymmetries are
excluded.

This suggests that, in order to improve the consistency of the 
precision electroweak data either one must postulate an error in the 
experimental determination of $A_{FB}^b$ and invoke new physics to 
raise the Higgs mass above the direct search limits 
\cite{Altarelli:2001wx}, or one must introduce new physics that 
directly modifies the couplings of the bottom quark in order to 
restore the consistency of the measurements \cite{Choudhury:2001hs}.

The RS model does not discriminate between the leptons and quarks, and 
thus this second option is denied to us.  Below, we will consider the
fits to precision electroweak parameters $S$, $T$, and $U$ both with 
and without $A_{FB}^b$.  While we do not advocate ignoring $A_{FB}^b$ 
in general, it is interesting to see that the RS model with bulk gauge 
bosons can in fact lead to a remarkable improvement of the description 
of the electroweak observables once the hadron asymmetries are 
ignored, for a Higgs mass $m_h$ well above the experimental limits.

We use the $S$,$T$, and $U$ fits of \cite{Altarelli:2001wx}, which,
for $m_t = 174.3$ GeV yield\footnote{These fits do not differ very 
substantially from i.e., those obtained in \cite{Langacker:2002sy}.} 
(removing $A_{FB}^b$ from the fit),
\bea
\label{eq:stulep}
S & = & -0.14 \pm 0.12
\nonumber\\
T & = & -0.08 \pm 0.13
\nonumber\\
U & = & ~~0.20 \pm 0.14 ~,
\eea
or (keeping $A_{FB}^b$ in the fit),
\bea
\label{eq:stuhad}
S & = & ~~0.00 \pm 0.11
\nonumber\\
T & = & -0.03 \pm 0.13
\nonumber\\
U & = & ~~0.27 \pm 0.14 ~.
\eea
We compute the full $S_{\rm{eff}}$, $T_{\rm{eff}}$, and $U_{\rm{eff}}$ as
the sum of the extra dimensional
contributions, Eqs.~(\ref{Seff})--(\ref{Ueff}), and also
the contributions from the Higgs \cite{Peskin:1991sw},
\bea
S_H & \simeq & \frac{1}{12 \pi}
\log \left( \frac{m_h^2}{ {m^2_{ref}}} \right) \\
T_H & \simeq & -\frac{3}{16 \pi c_0^2}
\log \left( \frac{m_h^2}{ {m^2_{ref}}} \right) \\
U_H & \simeq & 0 ,
\eea
where $m_{ref}$ is a reference Higgs mass, chosen for the fits of
Eqs.~(\ref{eq:stulep}) and (\ref{eq:stuhad}) to be 113 GeV.

We scan the parameter space of $L$ and $r$ for two fixed values of 
$m_h = 115$ and $200$ GeV, and identify the regions of $r$ and $L$ 
which are consistent with Eq.~(\ref{eq:stulep}) or (\ref{eq:stuhad}) 
at $1 \sigma$, $2 \sigma$, and $3 \sigma$.  We define consistency at 
the $n \sigma$ level to mean that all three oblique parameters 
predicted by the RS model are within $n-\sigma$ error intervals of the 
central fitted values.  We fix the top quark mass to its experimental 
mean value, $m_t = 174.3$ GeV. The one-sigma variations of the top
quark mass lead to corrections of about $\Delta S_t \simeq \pm 0.01$, $\Delta
T_t \simeq \pm 0.06$ and $\Delta U_t \simeq \pm 0.025$, which are
smaller than the one $\sigma$ errors on $S$, $T$ and $U$ obtained from 
the fit to the data and do not affect our results in a relevant way.

When comparing how well RS fits the data compared to the SM, it is 
important to remember that the SM fails this analysis for any region 
of parameters at $1 \sigma$, and agrees at roughly $2 \sigma$ for 
Higgs masses above the LEP limit and below about 200 GeV. Furthermore,
for the same range of Higgs mass parameters, removing $A_{FB}^b$
does not improve the agreement between the experimental data and the standard
model predictions.

\begin{figure}[t]
\vspace*{-1.cm}
\centerline{ \hspace*{1cm}
\epsfxsize=10.0cm\epsfysize=10.0cm
		     \epsfbox{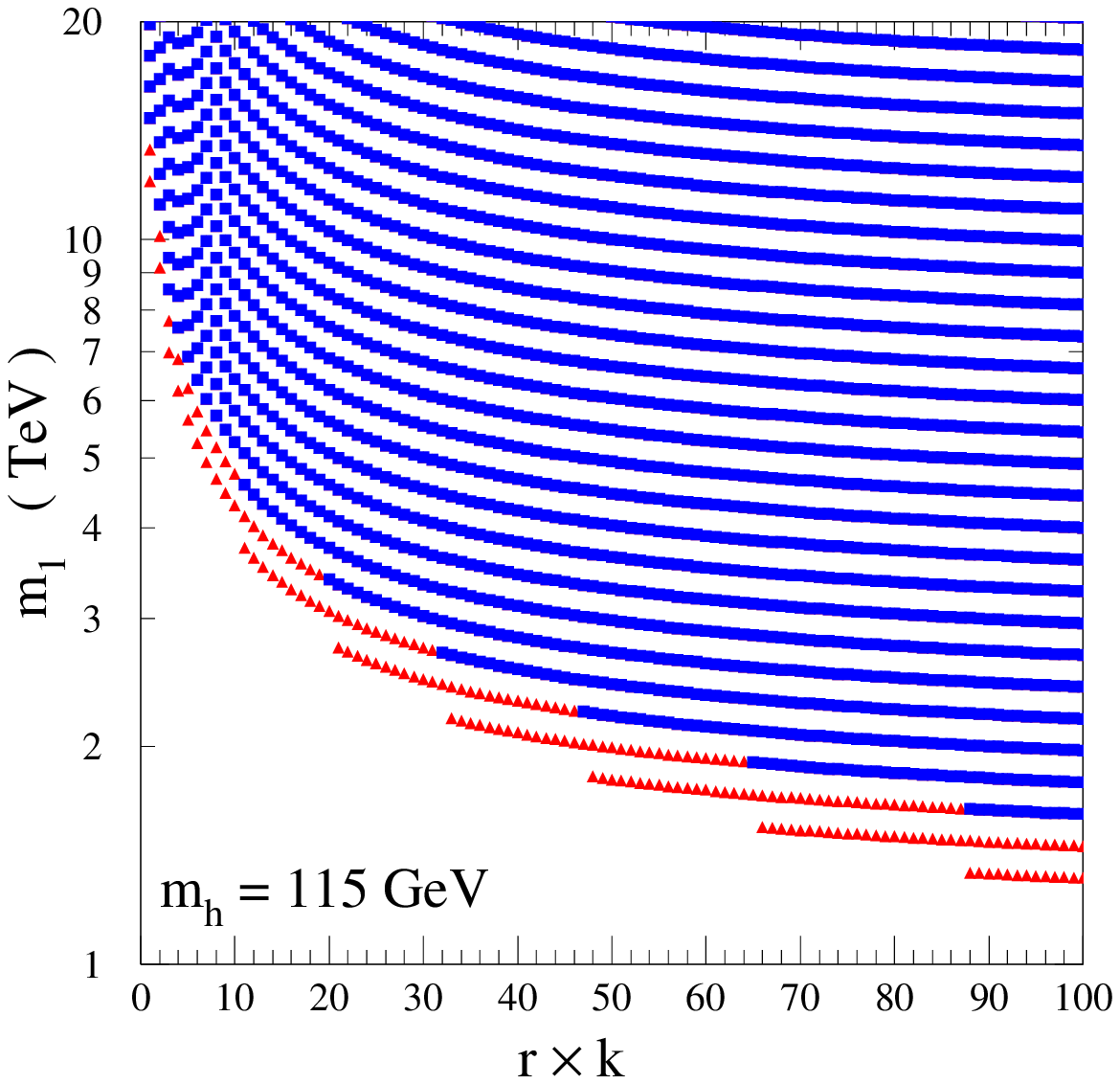} \hspace*{-1.5cm}
\epsfxsize=10.0cm\epsfysize=10.0cm
		     \epsfbox{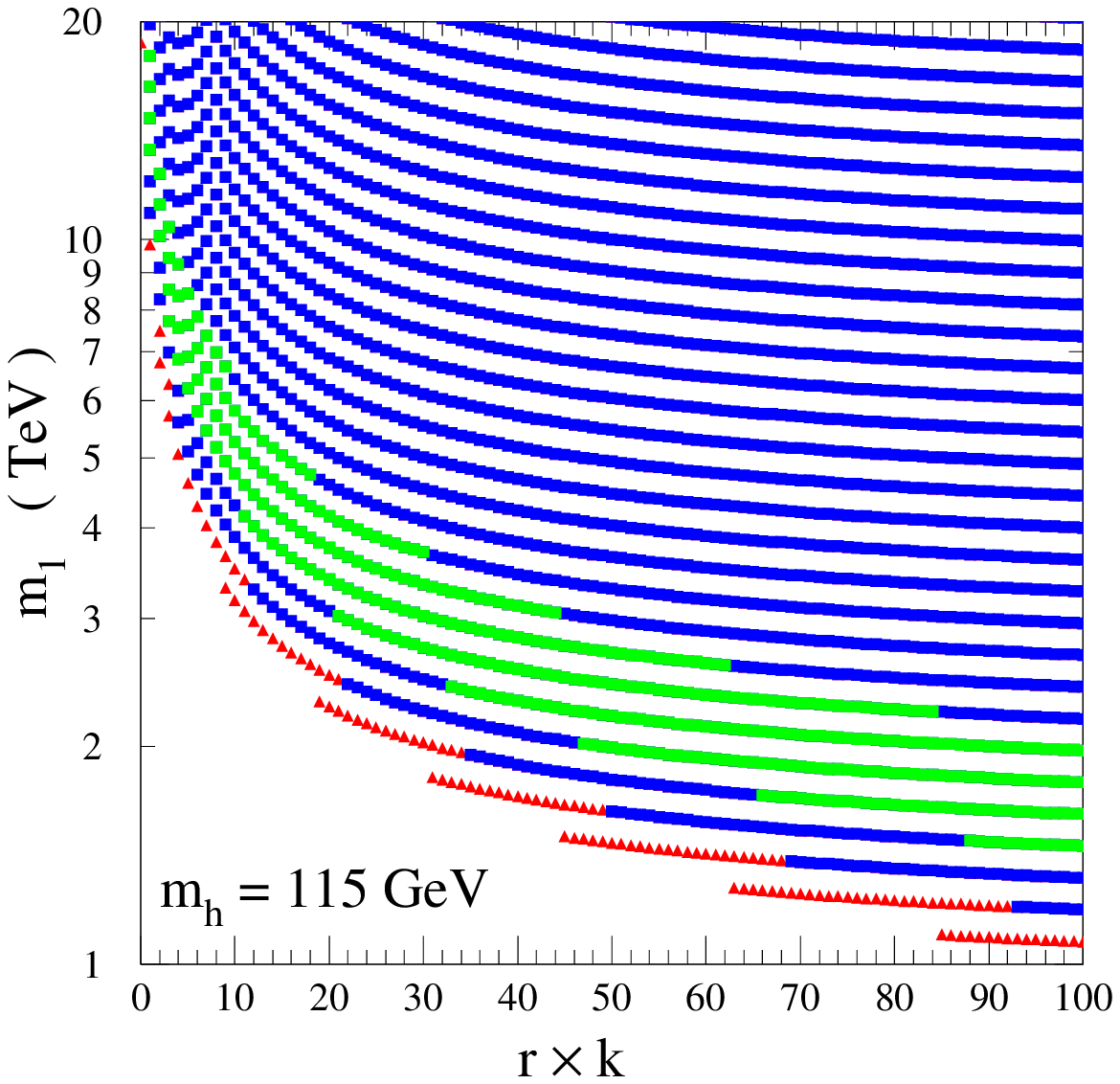}
}
\caption{The regions of $1 \sigma$ [green (light grey) squares],
$2 \sigma$ [blue (dark grey) squares]
and $3 \sigma$ [red triangles] agreement with precision electroweak
observables as defined in the text,
in the plane of $r \times k$ and the mass of the first KK
photon.  The Higgs mass has been fixed to $m_h=115$ GeV.
The left figure includes all precision data in the fit, whereas
the right figure includes only leptonic determinations of
$\sin^2 \theta_W$.}
\label{fig:fitmh115}
\end{figure}

The results for $m_h = 115$ GeV are shown in Fig.~\ref{fig:fitmh115}, 
plotted in the plane of $r$ and $m_1$, the mass of the first KK mode 
of the photon.  Generally, the masses of the first KK modes of the $W$ 
and $Z$ bosons will be somewhat larger, because of the electroweak
symmetry-breaking effects.  In the limit
$r \ra 0$, we find agreement with the limit obtained in
Ref.  \cite{Csaki:2002gy}, $m_1 \gsim 27$ TeV. As $r$ increases, there
are two different effects.  For small non-zero $r$, the overall fit to 
the data remains roughly the same as the $r=0$ case (in other words, 
the properties of the zero modes do not change much), but the 
properties of the first KK mode and its coupling to fermions are quite 
radically affected, with both dropping.\footnote{Note that, for $v/k \ll 1$,
as demanded by the fit to the data, the properties of the KK mode masses
and couplings can be read from Fig. 1.}  This results in the KK photon
being much lighter than one would have assumed for zero $r$, and, for 
moderate values of $r$, its coupling is comparable to that of the zero 
mode photon.

For $r \times k \sim 20$ and $m_1$ on the order of 3.5 TeV, the fit to
the entire data set is roughly as good as the SM itself, and the fit 
to the data without the hadron asymmetries is actually consistent
within $1 \sigma$.  Indeed, for the large region of parameters denoted 
by light squares in Fig.~\ref{fig:fitmh115}, we obtain an improvement 
of the fit to the data with respect to the Standard Model, with $m_1$ 
of order of a few TeV.

\begin{figure}[t]
\vspace*{-1.cm}
\centerline{ \hspace*{1cm}
\epsfxsize=10.0cm\epsfysize=10.0cm
		     \epsfbox{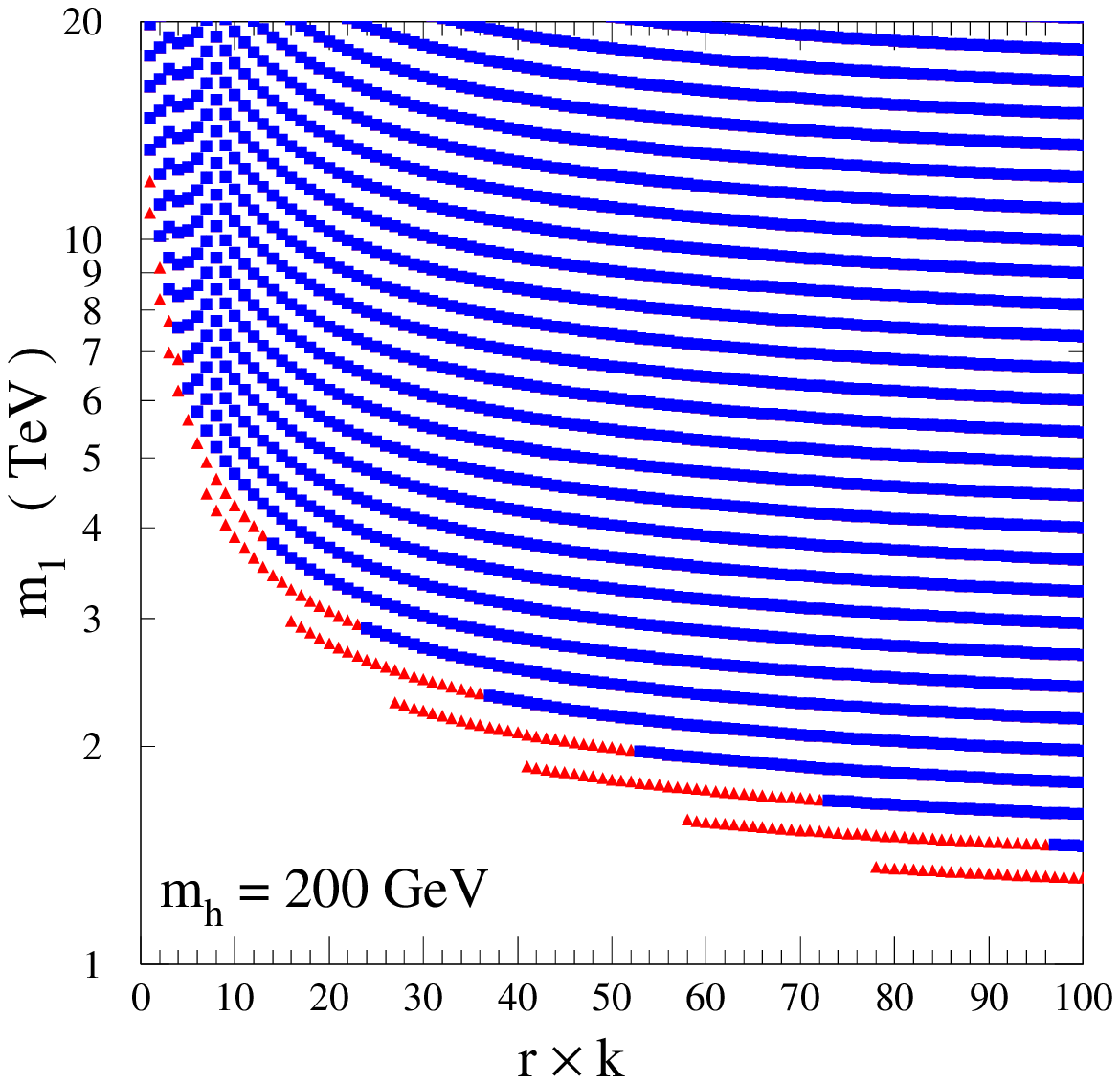} \hspace*{-1.5cm}
\epsfxsize=10.0cm\epsfysize=10.0cm
		     \epsfbox{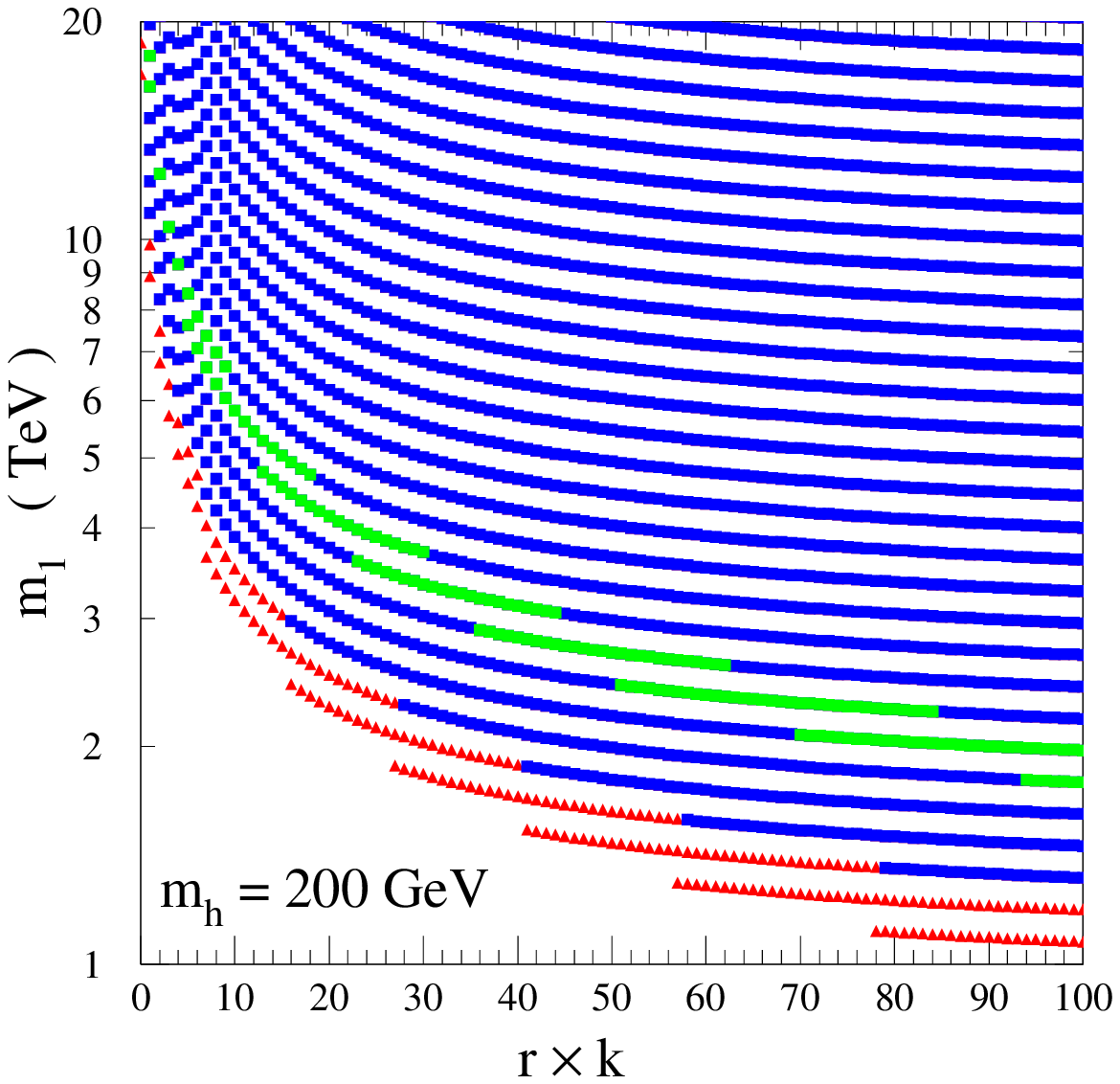}
}
\caption{The
same as Fig.~\ref{fig:fitmh115}, but for $m_h = 200$ GeV.}
\label{fig:fitmh200}
\end{figure}

In Fig.~\ref{fig:fitmh200} we show the same analysis, but with the 
Higgs mass fixed to $m_h = 200$ GeV. We see some small variation of 
the allowed regions, but by relatively small amounts, indicating that 
the RS fit does not prefer a light Higgs, but fits about equally well 
to Higgs masses as large as 200 GeV. The analysis for larger $m_h$ is 
straight-forward, and shows somewhat worse agreement than for $m_h = 
200$ GeV.

Our results have very important implications for the RS model with 
gauge fields in the bulk.  The preferred parameter space, for moderate 
$50 \simgt k\,r \simgt 10$, has a first KK photon with fermionic 
couplings of the order of the zero mode couplings, and a mass of a few 
TeV, most likely within reach of the LHC experiments.

\section{Conclusions}
\label{sec:concl}

The RS model is an interesting construction, with a novel solution to 
the hierarchy problem.  While the simplest versions have gauge fields 
confined to the IR brane, there are interesting motivations to allow 
them to propagate into the extra dimension, including grand 
unification and a relation to a class of walking technicolor theories 
through the ADS/CFT correspondance.  In that case, the lowest KK modes 
of the gauge bosons are usually very strongly coupled to fermions on 
the IR brane, with the lowest mass given by a factor of a few times $k 
e^{-kL}$, at the TeV scale.

Precision electroweak observables are challenging to fit in this 
framework, owing to corrections to the zero mode $W$ and $Z$ boson 
wave functions induced by the presence of the Higgs boson in the 
infra-red brane.  In the absence of local gauge kinetic terms, 
consistency of the fit to the precision electroweak observables at the 
2$\sigma$ level, as defined in the text, demands masses of the KK
modes of the gauge bosons larger than about 27 TeV, beyond the reach
of the next generation of collider experiments.

However, the inclusion of brane kinetic terms is helpful in lowering 
this bound.  For moderately large values of $r$, 2-$\sigma$
consistency may be obtained for KK gauge boson masses of the order of 
a few TeV. Furthermore, if the hadronic
asymmetries are excluded from the fit, the bulk gauge bosons may lead
to a consistent fit to the data at the 1 $\sigma$ level for a lightest 
KK mode of the order of a few TeV and Higgs masses in the range 
115-200 GeV. These light KK modes have couplings of order of the
zero mode coupling, opening new exciting possibilities at the LHC.

\newpage

~\\
{\Large \bf Acknowledgements}\\
~\\
The authors are grateful for conversations with G. Burdman.
Work at ANL is supported in part by the US DOE, Div.\ of HEP,
Contract W-31-109-ENG-38.
Fermilab is operated by Universities Research
Association Inc. under contract no. DE-AC02-76CH02000 with the DOE.

~\\
{\bf Note added:} Related work on this subject appeared at a
similar time in Ref.~\cite{Davoudiasl:2002ua}.
We have explored the case $r<0$ proposed in this
reference and we found that the bounds from $S_{\rm{eff}}$, $T_{\rm{eff}}$,
and $U_{\rm{eff}}$  are more constraining than
those based on the KK mode contributions to the muon decay constant alone.


\begin{thebibliography}{99}

\bibitem{Arkani-Hamed:1998rs}
N.~Arkani-Hamed, S.~Dimopoulos and G.~R.~Dvali,
Phys.\ Lett.\ B {\bf 429}, 263 (1998)
[arXiv:hep-ph/9803315];
I.~Antoniadis, N.~Arkani-Hamed, S.~Dimopoulos and G.~R.~Dvali,
Phys.\ Lett.\ B {\bf 436}, 257 (1998)
[arXiv:hep-ph/9804398].

\bibitem{Randall:1999ee}
L.~Randall and R.~Sundrum,
Phys.\ Rev.\ Lett.\  {\bf 83}, 3370 (1999)
[arXiv:hep-ph/9905221].

\bibitem{Davoudiasl:1999tf}
H.~Davoudiasl, J.~L.~Hewett and T.~G.~Rizzo,
Phys.\ Lett.\ B {\bf 473}, 43 (2000)
[arXiv:hep-ph/9911262].

\bibitem{Csaki:2002gy}
C.~Csaki, J.~Erlich and J.~Terning,
Phys.\ Rev.\ D {\bf 66}, 064021 (2002)
[arXiv:hep-ph/0203034].

\bibitem{Burdman:2002gr}
G.~Burdman,
Phys.\ Rev.\ D {\bf 66}, 076003 (2002)
[arXiv:hep-ph/0205329].

\bibitem{Huber:2000fh}
S.~J.~Huber and Q.~Shafi,
Phys.\ Rev.\ D {\bf 63}, 045010 (2001)
[arXiv:hep-ph/0005286];
S.~J.~Huber, C.~A.~Lee and Q.~Shafi,
Phys.\ Lett.\ B {\bf 531}, 112 (2002)
[arXiv:hep-ph/0111465].

\bibitem{Dvali:2000rx}
G.~R.~Dvali, G.~Gabadadze and M.~A.~Shifman,
Phys.\ Lett.\ B {\bf 497}, 271 (2001)
[arXiv:hep-th/0010071].

\bibitem{Georgi:2000ks}
H.~Georgi, A.~K.~Grant and G.~Hailu,
Phys.\ Lett.\ B {\bf 506}, 207 (2001)
[arXiv:hep-ph/0012379].

\bibitem{Carena:2002me}
M.~Carena, T.~M.~P.~Tait and C.~E.~M.~Wagner, Acta Physica Polonica, B33, 2355 (2002)
arXiv:hep-ph/0207056.

\bibitem{Carena:2001xy}
M.~Carena, A.~Delgado, J.~Lykken, S.~Pokorski, M.~Quiros and C.~E.~Wagner,
Nucl.\ Phys.\ B {\bf 609}, 499 (2001)
[arXiv:hep-ph/0102172].

\bibitem{Dvali:2001gm}
G.~R.~Dvali, G.~Gabadadze, M.~Kolanovic and F.~Nitti,
Phys.\ Rev.\ D {\bf 64}, 084004 (2001)
[arXiv:hep-ph/0102216].

\bibitem{gaugerunning}
L.~Randall and M.~D.~Schwartz,
JHEP {\bf 0111}, 003 (2001)
[arXiv:hep-th/0108114];
Phys.\ Rev.\ Lett.\  {\bf 88}, 081801 (2002)
[arXiv:hep-th/0108115];
K.~w.~Choi, H.~D.~Kim and Y.~W.~Kim,
arXiv:hep-ph/0202257;
K.~Agashe, A.~Delgado and R.~Sundrum,
Nucl.\ Phys.\ B {\bf 643}, 172 (2002)
[arXiv:hep-ph/0206099];
A.~Lewandowski, M.~J.~May and R.~Sundrum,
arXiv:hep-th/0209050.
K.~Agashe, A.~Delgado and R.~Sundrum,
arXiv:hep-ph/0212028.

\bibitem{Goldberger:2002cz}
W.~D.~Goldberger and I.~Z.~Rothstein,
arXiv:hep-th/0204160;
arXiv:hep-th/0208060.

\bibitem{Ponton:2001hq}
E.~Pont\'{o}n and E.~Poppitz,
JHEP {\bf 0106}, 019 (2001)
[arXiv:hep-ph/0105021].

\bibitem{Pomarol:1999ad}
A.~Pomarol,
Phys.\ Lett.\ B {\bf 486}, 153 (2000)
[arXiv:hep-ph/9911294].

\bibitem{Peskin:1991sw}
M.~E.~Peskin and T.~Takeuchi,
Phys.\ Rev.\ D {\bf 46}, 381 (1992).

\bibitem{Rizzo:1999br}
T.~G.~Rizzo and J.~D.~Wells,
Phys.\ Rev.\ D {\bf 61}, 016007 (2000)
[arXiv:hep-ph/9906234].

\bibitem{LEPTCG}
http://lepewwg.web.cern.ch/LEPEWWG/lepww/tgc/

\bibitem{Arkani-Hamed:2000ds}
N.~Arkani-Hamed, M.~Porrati and L.~Randall,
JHEP {\bf 0108}, 017 (2001)
[arXiv:hep-th/0012148].

\bibitem{Kennedy:1988sn}
D.~C.~Kennedy and B.~W.~Lynn,
Nucl.\ Phys.\ B {\bf 322}, 1 (1989).

\bibitem{Altarelli:2001wx}
G.~Altarelli, F.~Caravaglios, G.~F.~Giudice, P.~Gambino and G.~Ridolfi,
JHEP {\bf 0106}, 018 (2001)
[arXiv:hep-ph/0106029].

\bibitem{Csaki:2000zn}
C.~Csaki, M.~L.~Graesser and G.~D.~Kribs,
Phys.\ Rev.\ D {\bf 63}, 065002 (2001)
[arXiv:hep-th/0008151].

\bibitem{Giudice:2000av}
G.~F.~Giudice, R.~Rattazzi and J.~D.~Wells,
Nucl.\ Phys.\ B {\bf 595}, 250 (2001)
[arXiv:hep-ph/0002178].

\bibitem{Chanowitz:2001bv}
M.~S.~Chanowitz,
Phys.\ Rev.\ Lett.\  {\bf 87}, 231802 (2001)
[arXiv:hep-ph/0104024].

\bibitem{Choudhury:2001hs}
D.~Choudhury, T.~M.~P.~Tait and C.~E.~M.~Wagner,
Phys.\ Rev.\ D {\bf 65}, 053002 (2002)
[arXiv:hep-ph/0109097].

\bibitem{Langacker:2002sy}
P.~Langacker,
arXiv:hep-ph/0211065;
LEP Electroweak Working Group, http://lephiggs.web.cern.ch/LEPEWWG/.

\bibitem{Davoudiasl:2002ua}
H.~Davoudiasl, J.~L.~Hewett and T.~G.~Rizzo,
arXiv:hep-ph/0212279.

\end{thebibliography}
\end{document}